%% file: main.tex
\newif\iffull
\newif\ifsubmission \submissiontrue 
\newif \ifarxiv \arxivtrue
\documentclass[sigconf, nonacm]{acmart}

\usepackage[utf8]{inputenc}
\usepackage{xspace}
\usepackage{hyperref}
\usepackage{url}

\usepackage{breakurl}
\usepackage{enumitem}
\usepackage{booktabs}
\usepackage{multirow}
\usepackage{balance}
\usepackage{csvsimple}
\usepackage{array, makecell}
\usepackage[
  separate-uncertainty = true,
  multi-part-units = repeat
]{siunitx}
\usepackage{mhchem}
\usepackage{cleveref}
\dblfloatsep=\floatsep
\dbltextfloatsep=\textfloatsep
\usepackage{amsthm}
\theoremstyle{definition}

\usepackage{tabularx}
\usepackage{makecell}
\usepackage{array}
\usepackage{graphicx}
\graphicspath{{./Images/}}
\usepackage{xstring}
\usepackage{xcolor}
\usepackage{supertabular}
\usepackage{longtable}
\newcommand{\ethaddr}[1]{\textcolor{blue}{\href{https://etherscan.io/address/#1}{\StrLeft{#1}{4}...}}}%...\StrRight{#1}{4}
\newcommand{\trxaddr}[1]{\textcolor{blue}{\href{https://tronscan.org/\#/address/#1}{\StrLeft{#1}{4}...}}}%...\StrRight{#1}{4}
\newif \ifcomments \commentstrue
\ifcomments
    \newcommand{\ari}[1]{\textsf{\color{blue}{[Ari: {#1}]}}}
    \newcommand{\sarah}[1]{\textsf{\color{violet}{[SarahA: {#1}]}}}
    \newcommand{\sarahm}[1]{\textsf{\color{red}{[Sarah M: {#1}]}}}
    \newcommand{\h}[1]{\textsf{\color{purple}{[H: {#1}]}}}
    \newcommand{\tyler}[1]{\textsf{\color{cyan}{[Tyler: {#1}]}}}
    \newcommand{\taskassign}[1]{\textsf{\color{red}{[TASK: {#1}]}}}
    \newcommand{\todo}[1]{\begin{large}{\textbf{\color{red}{ [TODO: {#1}] }}}\end{large}}
\else
    \newcommand{\ari}[1]{}
    \newcommand{\sarah}[1]{}
    \newcommand{\h}[1]{}
    \newcommand{\tyler}[1]{}
    \newcommand{\taskassign}[1]{}
    \newcommand{\todo}[1]{}
\fi

\begin{document}
%\title{\large \bf Forsage: Anatomy of a Smart-Contract Pyramid Scheme}
\title{Forsage: Anatomy of a Smart-Contract Pyramid Scheme}
\date{}

\author{Tyler Kell}
\affiliation{Cornell Tech}
\affiliation{IC3}
\email{sk3259@cornell.edu}
\author{Haaroon Yousaf}
\affiliation{University College London}
\affiliation{IC3}
%\email{h.yousaf@ucl.ac.uk}
\author{Sarah Allen}
\affiliation{Cornell Tech}
\affiliation{IC3}
%\email{sarahallen@cornell.edu}
\author{Sarah Meiklejohn}
\affiliation{University College London}
\affiliation{IC3}
%\email{s.meiklejohn@ucl.ac.uk}
\author{Ari Juels}
\affiliation{Cornell Tech}
\affiliation{IC3}
%\email{juels@cornell.edu}

% Abstract
\begin{abstract}

Pyramid schemes are investment scams in which top-level participants in a hierarchical network recruit and profit from an expanding base of defrauded newer participants. Pyramid schemes have existed for over a century, but there have been no in-depth studies of their dynamics and communities because of the opacity of participants’ transactions. 
 
In this paper, we present an empirical study of Forsage, a pyramid scheme implemented  as a smart contract and at its peak one of the largest consumers of resources in Ethereum. As a smart contract, Forsage makes its (byte)code and all of its transactions visible on the blockchain. We take advantage of this unprecedented transparency to gain insight into the mechanics, impact on participants, and evolution of Forsage.

We quantify the (multi-million-dollar) gains of top-level participants as well as the losses of the vast majority (around 88\%) of users. We analyze Forsage code both manually and using a purpose-built transaction simulator to uncover the complex mechanics of the scheme. Through complementary study of promotional videos and social media, we show how Forsage promoters have leveraged the unique features of smart contracts to lure users with false claims of trustworthiness and profitability, and how Forsage activity is concentrated within a small number of national communities.

\end{abstract}

\maketitle

\input{Sections/all}

\bibliographystyle{ACM-Reference-Format}
\bibliography{references}

\appendix
\input{Sections/appendixA}

\input{Sections/appendixB}
\input{Sections/appendixC}
\input{Sections/appendixD}

\end{document}

%% file: Sections/all.tex
%!TEX root = ../main.tex

% Sections
\input{Sections/introduction}
\input{Sections/background}

\input{Sections/overview}
\input{Sections/evaluationSmartContract}
\input{Sections/statistics}

\input{Sections/community}
\input{Sections/howToFix}
\input{Sections/relatedWork}

\input{Sections/conclusion}
\input{Sections/acknowledgments}

%% file: Sections/introduction.tex
\section{Introduction}

Cryptocurrencies and smart contracts are new and powerful technologies that promise a range of benefits, including faster monetary transactions, innovative financial instruments, and global financial inclusion for the world’s unbanked. Conversely, though, these same technologies have fueled new forms of fraud and theft~\cite{zhao02017, Ferreira2019} and new ways of perpetrating existing types of crime~\cite{ransomwarejusticedept,phillips2020tracing}. %\sarahm{Someone needs to be in charge of filling in these citations.}\sarah{on it}
 
{\em Pyramid schemes}, for example, are a prevalent type of scam in which top-tier participants in a hierarchical network recruit and profit at the expense of an expanding base of new participants. They have existed for more than a century, but have recently emerged in a new form: as smart contracts on blockchains such as Ethereum. 

Smart contracts are in some ways an ideal medium for pyramid schemes and other scams. Because they run in decentralized systems, they cannot easily be dismantled by law enforcement agencies. They can instantaneously ingest payments from victims across the globe. They provide privacy protection for their creators in the form of pseudonymous addresses. Finally, as so-called ``trustless’’ applications---with world-readable (byte)code---they present a veneer of trustworthiness to unsuspecting users. 

The flip side of such transparency is that smart contracts offer researchers a degree of visibility into the mechanics of online (and offline) scams that is without historical precedent. Not only is the (byte)code specifying the scam’s mechanics visible on chain, but so is every transaction performed by every participant. 

In this paper, we take advantage of this newfound visibility to conduct an in-depth measurement study of the largest smart contract-based pyramid scheme to date, called {\em Forsage Smartway} or {\em Forsage} for short. 

Forsage came into existence in late January 2020. It was at one point the second most active contract in Ethereum by daily transaction count, and remains in the top twenty at the time of writing.  As we show throughout this paper, it is a classic pyramid scheme, defined by the SEC as ``a type of fraud in which participants profit almost exclusively through recruiting other people to participate in the program''~\cite{sec.gov_2013}. The Forsage contract requires players to send currency (Ether) in order to participate. Funds sent by newly recruited users immediately pass through the contract to existing players, with those at the top of the (smart contract-defined) pyramid obtaining the largest returns. 

Understanding the success of Forsage requires study of not just the contract itself, but also its community of hundreds of thousands of users, many of whom have actively discussed and marketed the scam. Consequently, to paint a detailed picture of how Forsage lures and defrauds users, our study combines measurement and analysis of a range of complementary forms of data, including source code, on-chain transaction data, and social media interactions.

\subsection{Main study results}

Our results come from three basic, mutually illuminating forms of study: smart contract deconstruction (Section~\ref{sec:evaluation}), blockchain analytics (Section~\ref{sec:measurement_study}), and analysis of video and social media interactions (Section~\ref{sec:community}).

We believe that our study's findings are not just relevant to Forsage, but provide durable insights into the conception, mechanics, and evolution of smart-contract scams and financial scams more generally. They also point to effective strategies that  
government authorities and the cryptocurrency community can use to combat pyramid schemes and other scams, as we discuss in Section~\ref{sec:solutions}.

\paragraph{Contract deconstruction:} Forsage promoters highlight the ``transparency'' of the contract, emphasizing that its source code is public.  

We present the results of our effort to deconstruct Forsage and obtain a detailed description of the scheme's dynamics.  We show that the scheme does indeed operate as a pyramid, with a small number of users at the top after having recruited a much larger number of users.  

We also quantify the cost of Forsage's complexity in terms of on-chain transaction fees, showing that Forsage transactions consume far more gas than most other Ethereum transactions and thus cost more for its users. 

\paragraph{Contract measurement study:}

Through a measurement of Forsage transactions on the Ethereum blockchain, we document the flow of 721k ETH (226M USD) through Forsage from its creation to January 2021. 
One of our most striking findings is characteristic of pyramid schemes: The vast majority of Forsage players have lost money, with net losses for over 88\% of players.  A small few at the top of the pyramid have profited handsomely, e.g., the contract owner, who has received over 5000~ETH (1.2M~USD). To the best of our knowledge, our study offers the first precise quantification of payouts and losses in any large pyramid scheme, internet-based or historical.

\paragraph{Community study:}

Through analysis of videos, social media, and contract evolution, we characterize the dynamics of the Forsage community.

False claims of profitability and trustworthiness are typical of pyramid schemes and well represented in Forsage marketing videos and social media posts. We manually coded the repeated claims made by the top English-language Forsage videos---for example, the false claim that players can earn thousands of dollars without recruiting others. Forsage videos additionally underscore the ways in which Forsage innovates on previous pyramid schemes, however, by leveraging smart contracts, for instance, with the claim that smart contracts {\em inherently cannot be scams.}

Our study also sheds light on the background and evolution of the Forsage community. Using a combination of location data from Facebook, tagged tweets, and YouTube channel annotations, we show that Forsage activity is internationally broad, but highly concentrated within a few geographies (e.g., western Africa).

\subsection{Summary of contributions}

In summary, the main contributions of our study of Forsage in this paper are:

\begin{itemize}
    \item {\em Contract deconstruction:} Using a tool for transaction simulation that is of possible independent interest, we detail the operating rules of Forsage and show the concentration of power and wealth at the top of its defined pyramid(s). 

    \item {\em Contract measurement study:} In a measurement study of Forsage contract activity on Ethereum, we document the flow of funds and show monetary losses by the vast majority of users. 

    \item {\em Community-dynamics study:} By tagging claims in promotional videos and studying social media interactions around Forsage, we document tactics used to attract users and the geographical distribution of users.

\end{itemize}

We emphasize that our results, which reveal a combination of classic and smart contract-specific scam characteristics, offer insights not just into Forsage, but into both blockchain and non-blockchain scams more generally. 

%% file: Sections/background.tex
\section{Background}

\subsection{Smart contracts}

Forsage is realized as a {\em smart contract}. Smart contracts are applications that execute on {\em blockchains}, decentralized systems that indelibly and immutably record transactions in an authoritative sequence and are best known as the platforms that realize cryptocurrencies such as Bitcoin.   

The most popular public (permissionless) blockchain for smart contracts today is {\em Ethereum}~\cite{buterin_buterin_2014}, whose native currency is known as {\em Ether} (ETH). Ethereum smart contracts are launched in the form of bytecode that runs in a Turing-complete environment known as the Ethereum Virtual Machine (EVM).  Smart contract creators often also publish corresponding source code, typically written in the Solidity programming language, but such publication is optional. {\em Transactions} sent to smart contracts by users are processed by contract code and are publicly visible on chain. 

Transactions may send money to a contract from user accounts or other contracts and must specify payment of execution fees to miners in the form of {\em gas}, a parallel currency converted into ETH upon transaction execution. 
This conversion is calculated by multiplying the amount of work performed by a transaction (its ``gas consumed") by the price of gas in ETH set by user when submitting the transaction~\cite{yellowpaper}.

Correctness of contract execution is enforced by the consensus mechanism underlying the Ethereum blockchain, so a miner's execution of contract code in the EVM must be agreed upon by all network participants to be included in a confirmed block. 

Other permissionless blockchains with smart contract functionality are growing in popularity, e.g., Tron~\cite{tronwhitepaper}, to which Forsage has also been ported. Ethereum, however, remains the dominant smart contract platform. 

\subsection{Scams}

Scams, i.e., fraudulent schemes involving financial deception, have been documented for centuries. Many scams involving large populations of victims assume the form of {\em pyramid schemes}. The U.S. Securities and Exchange Commission (SEC) defines a pyramid scheme as ``a type of fraud in which participants profit almost exclusively through recruiting other people to participate in the program''~\cite{sec.gov_2013}. Pyramid schemes, which are illegal in most jurisdictions, come in a number of variants. One variant is a {\em Ponzi scheme}, which specifically involves investment in financial instruments. {\em Multi-level marketing} (MLM) schemes, which involve the sale of a product or service, are related to pyramid schemes. They are legal in the U.S., but outlawed in some jurisdictions (e.g., China)~\cite{chinalawMLM}.

\subsection{Blockchain scams}
A multitude of scams have arisen within the blockchain ecosystem. Some scams have solicited investments from victims in new blockchain technologies. Examples include Onecoin, a Ponzi scheme that involved a fake (centralized) blockchain in which victims invested \$19+ billion ~\cite{madeira_2020},  Bitconnect, a token that promised returns of 1\% per day and saw investment of \$3.5 billion from victims, as well as other, related \$1+ billion schemes such as Plustoken and WoToken.pro ~\cite{palmer_2020, bel_2020}.

Other scams instead use blockchain technology to realize variants of scams, such as pyramid schemes, that were seen well before the advent of blockchains. Prominent examples are Million.Money\footnote{\url{https://million.money}} and Doubleway.io\footnote{\url{https://doubleway.io/}}, which are both currently active, as well as the defunct scheme Bullrun.live.\footnote{\url{http://bullrun.live}}  All three have similarities with Forsage: they use similar promotional materials, have a similar structure for the user dashboard, and use similar language and terminology (e.g., a referrer to the program is called an ``upline''). We explore Forsage user interactions with multiple scam contracts in ~\cref{subsec:user_behavior}.

%% file: Sections/overview.tex
\section{Forsage Overview}
\label{sec:overview}

The creators and promoters of Forsage advertise it as a \emph{matrix} MLM scheme, despite the lack of a service or product.  It operates primarily on Ethereum, where its initial Matrix contract has been active since January 31st, 2020. Since then, Forsage creators have also launched a Forsage contract on Tron (TRX) and an additional, followup smart contracts called Forsage xGold on both Tron and Ethereum. At the time of this writing, the Forsage authors are in the process of writing a Forsage Binance Smart Chain (BSC) contract.

\noindent \paragraph{The Forsage website:}
Users interact with Forsage using the forsage.io website, which shows how much they have paid into and earned from the contract. 
The website encourages the use of user-friendly cryptocurrency tools. It shows users how to purchase cryptocurrency using Trust Wallet, a user-friendly tool to exchange fiat for cryptocurrency, and how to use MetaMask, a browser extension that allows users to easily transact with cryptocurrency.  

The combination of these tools makes Forsage accessible to novice users who may not previously have used cryptocurrencies or smart contracts.  Screenshots of the Forsage website, showing the different matrices and their structure, can be found in  Appendix~\ref{sec:webpagescreenshots}.

\noindent\paragraph{Forsage use and structure:}

A new Forsage user must pay a minimum of 0.05 ETH, which opens up the \emph{slot} at the first \emph{level} in the two matrix systems, called X3 and X4. 
Each matrix consists of 12 slots. To unlock the ability to use the next slot (at level $i+1$), a user must pay twice as much ETH as for their currently highest slot (at level $i$). In both X3 and X4, the first slot costs 0.025~ETH, while the twelfth and final slot costs 51.2~ETH. This means that the total cost to open all slots in either matrix is 102.375~ETH.

Each Forsage user has a \emph{referral code}, created at the time they register. The referral code links a recruited user's account to the account that recruited them, called their \emph{upline}. These referral codes thus organize Forsage users into pyramids, with the oldest accounts at the top. Payments flow upwards within a pyramid as additional users join it. 

The pyramids of users linked by chains of referral code 
are referred to as Forsage \emph{teams}.  It is possible to join Forsage without entering a referral code; users who do so are assigned the referral code of the contract \emph{owner} (the creator).  

\Cref{sec:evaluation} contains an explanation of the logic for payment flow of user funds sent through the Forsage contract.  Briefly, users earn money in the X3 and X4 matrices as follows.

\begin{description}
    \item[X3:] In X3, users earn income by recruiting others into the system. A user must recruit three additional users to recoup their initial investment within each slot.
    Any recruits beyond the first three per slot will generate income for the recruiting user and those further up in their pyramid. Each subsequent slot costs more to open, but its resulting payout if filled with recruits will be higher because the expected payout for each three recruits is equal to the initial cost to open the slot for the recruiter. After a user fills a slot (i.e. recruits 3 users into that slot), Forsage \emph{blocks} the filled slot, causing the user to forfeit future earnings from it until it is unblocked. Unblocking means paying to open the slot at the next level up in the system, at which point this lower-level slot cannot become blocked again.
    
    \item[X4:]  \label{x4reference}  In X4, users can earn both by recruiting other users and by being on an active team. 
    When a user recruits the six additional users necessary to recoup their initial investment in an X4 slot (twice as many as are required in X3), that slot becomes blocked and the user will have received the same amount of money paid to open the slot, with others in their team getting paid as well. X4 also has an element of competition: If a newer user on a team is more active than the user whose referral code they used to join Forsage, that user can switch spots on the team, giving the more active, newer user the profits that would otherwise flow to the older, referring account~\cite{youtube1}. 

\end{description}

%% file: Sections/evaluationSmartContract.tex
\section{Forsage Contract Deconstruction}
\label{sec:evaluation}

Forsage promotional materials imply that the system is trustworthy because its code is open-source, e.g., the promotional materials claim that the contract ``guarantees the purity of conditions.'' 
We took advantage of the availability of the source code to conduct an in-depth analysis of the smart contract's logic and data structures.

\begin{figure}
  \centering
  %\vspace*{-2.1cm}
    \includegraphics[width=0.95\linewidth]{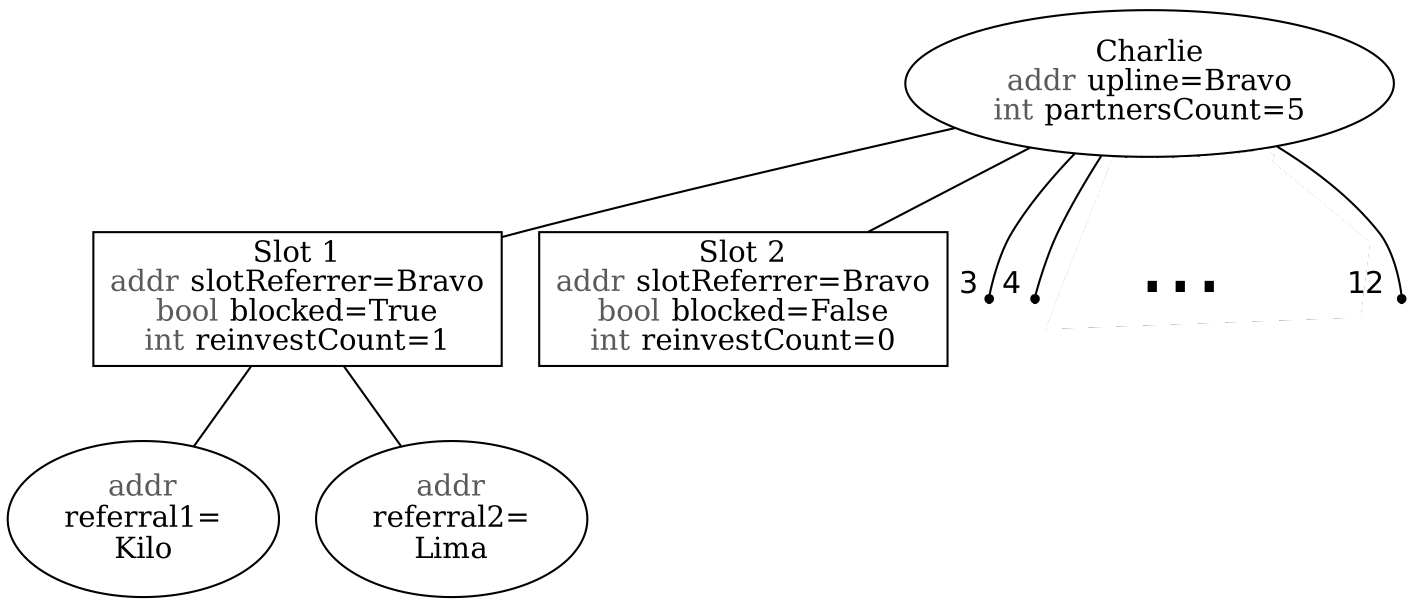}
    \caption{A visualization of the state the contract keeps for each user in the X3 matrix, focusing on a user Charlie. The \texttt{addr} variables point to Ethereum addresses, here given NATO-phonetic names. Matrix slots that have not yet been opened are depicted with a numbered dot, instead of a box.}
    \label{fig:data_structuresx3}
\end{figure}

\paragraph{Methodology and data collection:}
\label{para:data_collection}
The code for the Matrix smart contract is published on Etherscan.\footnote{ \fontsize{7.7}{12}\selectfont{\href{https://etherscan.io/address/0x5acc84a3e955Bdd76467d3348077d003f00fFB97}{etherscan.io/address/0x5acc84a3e955Bdd76467d3348077d003f00fFB97}}}
We first attempted manual source code review, but found the logic too confusing to follow without visualization. We then built a simulator in Python that deployed the contract to a local private test network of Go-Ethereum (Geth) nodes,\footnote{\url{https://github.com/ethereum/go-ethereum}} and used Web3.py\footnote{\url{https://github.com/ethereum/web3.py}} to send sample transactions. We also wrote a visualizer for the contract's state machine using GraphViz~\cite{Ellson01graphviz}. The output of that visualizer assisted in creating Figure~\ref{fig:data_structuresx3}, which depicts the data stored in the contract. Although the open source code is pointed to as a source of legitimacy by Forsage promotional materials, our analysis of the contract took weeks of focused effort by a professional research engineer. Our code for the visualizer and simulator tools are released as open source software on \href{GitHub}{https://github.com/initc3/forsage}.

When the Forsage team launched their Tron implementation of the Matrix smart contract, they also released its source code. We found this Tron code to be nearly identical to the Ethereum original, so we did not specifically analyze it. The latest iteration of Forsage launched on both Ethereum and Tron (as of May 2021), the xGold contract, has no publicly available source code.

The Ethereum and Tron blockchains include the data for all transactions performed by Forsage users.  We mined this publicly available data to perform further analysis.  
To obtain Ethereum data we ran the Go-Ethereum 
(Geth)\footnote{{\url{https://github.com/ethereum/go-ethereum}}} and TurboGeth\footnote{{\url{https://github.com/ledgerwatch/turbo-geth}}} full-node and archive-node software packages, and downloaded 
the entire blockchain up to January 14, 2021.

We then used the Ethereum-ETL\footnote{{\url{https://github.com/blockchain-etl/ethereum-etl}}} package
to retrieve this data from Geth and store the %343,135,303
345 million transactions included in the Ethereum blockchain between the launch of Forsage (January 31, 2020) and January 14, 2021.  We wrote custom Python scripts to analyze this data and found 222,516,680 transactions that involved function calls on smart contracts, of which 3,266,722  were to the Forsage smart contract. 
To profile user transactions
outside Forsage, we used the Chainalysis Reactor tool.\footnote{\url{https://www.chainalysis.com/chainalysis-reactor/}}  
Chainalysis Reactor is a web-based investigation platform that connects cryptocurrency transactions to real-world entities, using tags that are either internal to Chainalysis or gathered from public websites and documents.

To collect Tron transaction data we scraped the TronScan 
API\footnote{{\url{https://tronscan.org/}}} and parsed the results 
directly into CSV form.

\goodbreak

\paragraph{Forsage data structures:}
As discussed in Section~\ref{x4reference}, Forsage consists of two matrix systems, X3 and X4, each consisting of 12 slots.  These two matrices differ in the number of users that act to fill each matrix level (three for X3, six for X4) and the logic for how nodes propagate through them over time. 

The data for each user is stored in a hashtable (Solidity mapping) on the Ethereum blockchain, with the key being the user's address and the value being a Solidity \texttt{struct} with the data for that user's state tree and arrays of pointers to its children.
Figure~\ref{fig:data_structuresx3} visualizes this mapping for a user's X3 tree, with some minor metadata variables omitted.  Each user also has an X4 tree, whose structure is largely similar.  As seen in this figure, each user has an \emph{upline}, which is the user that referred them to the contract. This is distinct from \textsf{slotReferrer}, a variable used per slot as part of the payment logic. The \textsf{slotReferrer} variable is initialized to the upline, but changes over time as users refer each other. The \textsf{reinvestCount} variable keeps track of the number of times a slot has been filled. In our example, Charlie has filled his first matrix slot once already (and then unblocked it by buying a slot at level 2), meaning he has referred $3 \times \mathsf{reinvestCount} + 2 = 5 = \mathsf{partnersCount}$ users.

\paragraph{External API:}
The contract exposes 15 functions to read its state, and two state-changing functions, \texttt{registrationExt} and \texttt{buyNewLevel}. The first registers new users and thus adds them to the contract state. The second changes contract state for an existing user to allow them to continue to gain money from new referrals.

The placement of new users in the contract state depends on the X3 and X4 slots for the user that referred them (their upline). The logic of the contract \emph{scrambles} positions in the upline's matrices and in the matrices of the upline's parent when an upline's slot becomes full, i.e. every time the upline refers a multiple of three users to a given X3 slot ($\mathsf{partnersCount} \bmod 3 = 0$), or a multiple of six users to a given X4 slot. The logic of scrambling leaf nodes in the pyramid depends on the state of the slot \texttt{referrer} variable for the affected matrix slot, as well as the \texttt{blocked} variable for that slot, and in the X4 system an additional \texttt{closedPart} variable for each slot. Scrambling the positions of the existing users in the system helps to make payments through Forsage (falsely!) appear more random. It benefits older users in the pyramid, as users are usually scrambled ``up'' the pyramid to become children of older users rather than newer ones. 

\begin{table}
\centering
\small
\begin{tabularx}{0.999\linewidth}{XXXXX}
\toprule
Opcode & \makecell{Avg num\\ per tx\\ (all)} & \makecell{Median\\ (all)} & \makecell{Avg num\\ per tx\\ (Forsage)} & \makecell{Median\\ (Forsage)} \\
\midrule

\normalsize{\texttt{SSTORE}}  & \makecell{4.54\\ $\pm$ 8.10} & \makecell{2} & \makecell{10.76\\ $\pm$ 9.57} & \makecell{6} \\
\\
\normalsize{\texttt{SLOAD}}   & \makecell{17.84\\ $\pm$ 51.6} & \makecell{7} & \makecell{36.86\\ $\pm$ 26.21} & \makecell{29} \\
\bottomrule
\end{tabularx}
\caption{Average number of instruction operations per transaction, with standard deviation, for both all transactions and only those that interact with Forsage. Due to the intensive computation required to process this data, this table covers only the thousand blocks between block heights 10,600,000 and 10,601,000 (Roughly 13:00-18:00 UTC on August 5th, 2020) rather than our larger dataset including all transactions from 2020. This smaller dataset still contains 188,920 transactions that interact with smart contracts, 5667 of which interact with Forsage.}
\label{table:number_instructions_per}
\end{table}

\paragraph{Transaction fees:}
\label{subsec:gas}
The fact that Forsage has so much persistent on-chain storage means that its users pay higher gas fees than the average for Ethereum contracts, due to the heavy usage of the (expensive) \texttt{SLOAD} and \texttt{SSTORE} opcodes.
These fees are higher even when comparing Forsage transactions only to other contract function calls in Ethereum (so in particular ignoring simple sends of ETH). In our collected dataset of Ethereum network transactions, we found that the mean transaction fee for all Ethereum transactions that interacted with a contract was 0.00632~ETH with a standard deviation of 0.0618~ETH and a median of 0.00257~ETH.  Forsage transactions paid a higher average transaction fee of 0.0116~ETH with a standard deviation of 0.0108~ETH and a median of 0.00883~ETH. Forsage users pay more than four times as much on average as other smart contract users.

The most gas-expensive EVM operations/opcodes are those that create a new contract (\texttt{CREATE}, \texttt{CREATE2}); store, change, and access data into persistent on-chain state (\texttt{SSTORE}, \texttt{SLOAD}), and call contract functions or send money to other users in the network (\texttt{CALL})~\cite{yellowpaper}. Every transaction that interacts with Forsage through its two main functions, \texttt{registrationExt} and \texttt{buyNewLevel}, uses two of these three most expensive categories, often multiple times: they make use of persistent storage via \texttt{SSTORE} and \texttt{SLOAD} operations, and send money to other users on the network using Solidity operations that compile to the \texttt{CALL} opcode.  Forsage uses an average number of \texttt{CALL} operations, but makes heavy use of \texttt{SSTORE} and \texttt{SLOAD}, as shown in Table~\ref{table:number_instructions_per}.

\begin{figure}
    \centering
    \includegraphics[width=\linewidth]{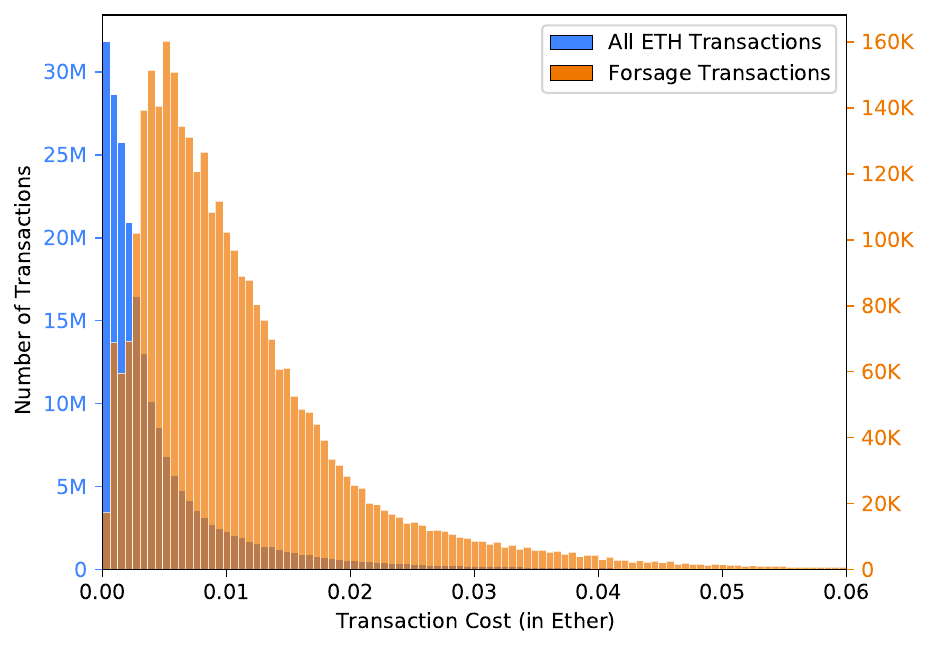}
    \caption{Histogram of transaction costs on the Ethereum blockchain---from January 31, 2020 to January 14, 2021---that involve successful smart contract function calls. Blue bars indicate the number of all transactions that paid fees within the given bucket, while orange bars indicate the same data, but only for transactions sent to the Forsage smart contract.  The data excludes outlier transactions with fees above 0.06~ETH, which is above the 99th percentile of all transactions from this time period.}
    \label{fig:tx_fee_histogram}
\end{figure}

Figure~\ref{fig:tx_fee_histogram} shows a superimposed histogram of Forsage transactions relative to all Ethereum transactions. 
The higher gas consumption associated with Forsage results in higher transaction fees overall, as demonstrated by the right-shifted peak in the Forsage curve relative to that of all ETH transactions.

\begin{figure}
\centering
\includegraphics[width=0.95\linewidth]{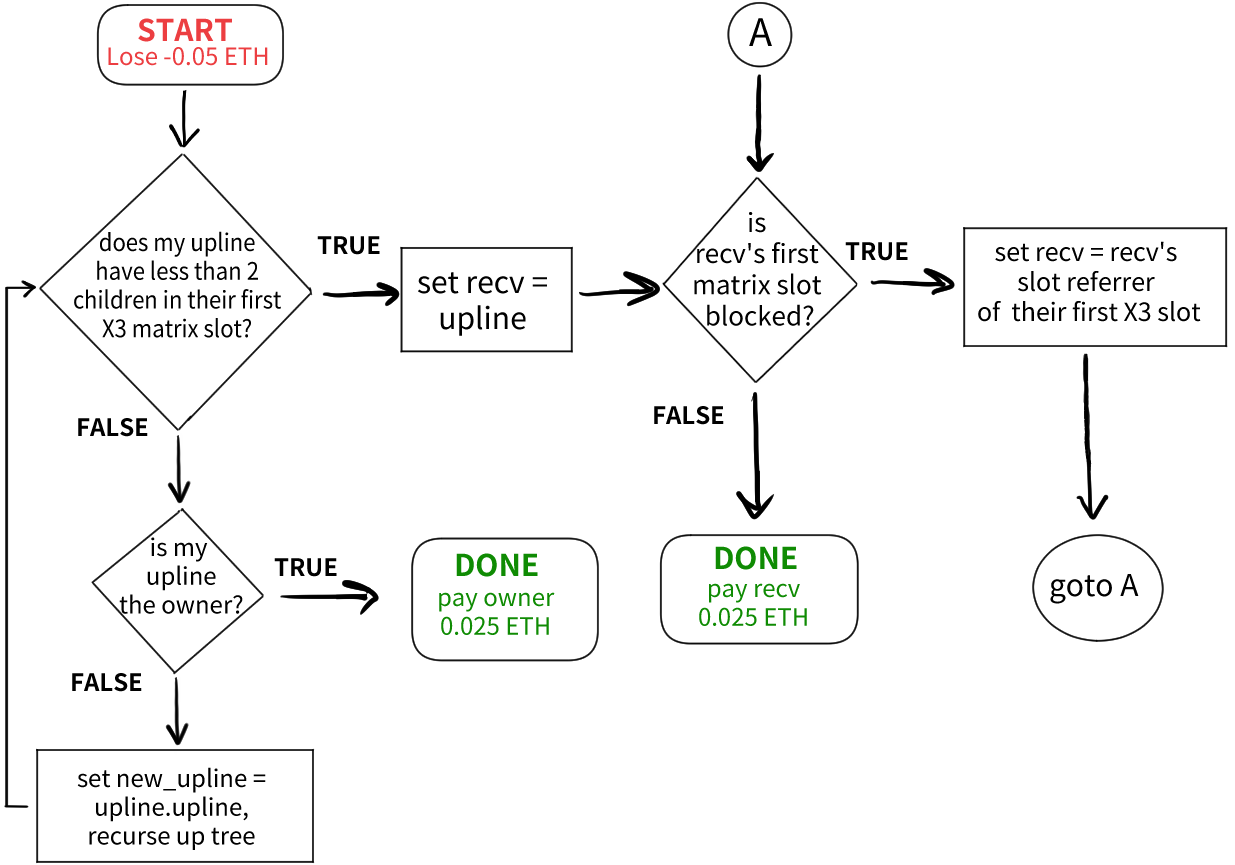}
\caption{Flow chart for the logic of who gets paid when a new user registers, in the X3 system. The \texttt{BuyNewLevel} function follows similar logic, but conditioned on the matrix slot being purchased, rather than the first slot.}
\label{forsage-payflow-x3}
\end{figure}

\paragraph{Payment logic:}

There are three ways for a user to get paid in Forsage: (1) by referring new users to the system; (2) when users they have referred in the past buy an additional matrix slot at a level corresponding to one previously purchased by the referrer; and (3) when \emph{spillover} occurs, a condition in the X4 matrix resulting from the slots of another user downstream in the pyramid being blocked. Whenever money is sent to the smart contract by one user, the contract atomically (i.e., in the same transaction) sends those funds to other users based on the logic described below. This allows Forsage promotional materials to claim that the contract ``never stores users' funds.''

When a user buys a new slot, the money they pay typically routes to the first found upline that also has that same slot open. Users are thus incentivized to buy new levels in order to refer users underneath them, which means a user can be generally successful by adding additional matrix slots just before referring additional users, and in general by recruiting as many users as possible.  

Figures~\ref{forsage-payflow-x3} and~\ref{forsage-payflow-x4} show the logic determining who gets paid when a new user registers with the Forsage contract, for both the X3 and X4 matrices. The logic for purchases of new slots (\texttt{buyNewLevel}) is largely similar but depends on the slot purchased rather than the first one (e.g., if a user buys the third slot then the logic is conditioned on the status of their upline's third slot).

\begin{figure}
    \centering
    \includegraphics[width=0.95\linewidth]{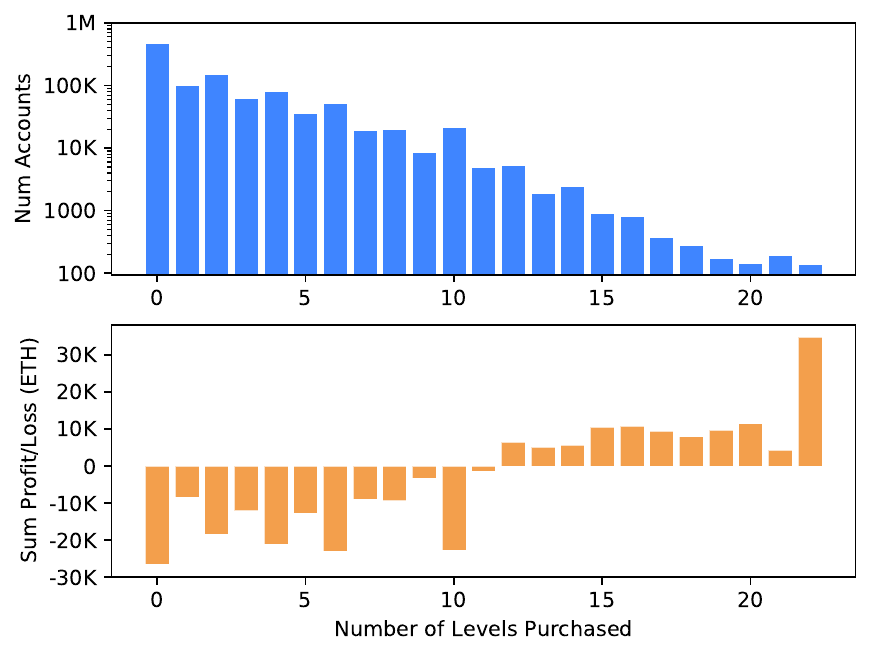}
    \caption{The distribution of how many users had unlocked a given number of levels in the contract (on top, and at log scale), and the collective amount of money gained or lost by the users who had unlocked this number of levels (on bottom, and at linear scale).  Users that bought the most levels were on average the most profitable.}
    \label{fig:num_levels_bought_histogram}
\end{figure}

The flowcharts in these figures show that uplines must keep their slots from becoming blocked, or payments will skip over them.  
To prevent a slot from becoming blocked, a user must buy the slot at the next level. This will also unblock an existing slot if it already has become blocked, and prevent the slot at $level - 1$ from ever becoming blocked again. Figure~\ref{fig:num_levels_bought_histogram} shows the distribution of levels purchased in aggregate for all users in the Forsage contract, as well as the summed profitability for the group of users that purchased that many slot levels. In general users that purchased more levels were also the most profitable users: The average user of the contract purchased 2.13 levels, with a standard deviation of 2.89 and a median of 1 level purchased.

When a new user joins the system, their payment is split into two equal parts and the logic in the flowchart is applied to each half, with one half going through the X3 flowchart and one half through the X4 flowchart, to determine which other user(s) should get each half of the payment.
If the direct upline of this new user is not blocked, then the upline gets the payment.  If the upline has a blocked slot, the contract checks the upline's upline for that matrix slot level to see if it is blocked. This iterates through uplines until the contract finds one that is unblocked, which it then pays.  The contract owner (i.e., the user that created the contract) is always unblocked, so the contract always finds a user to pay. This can sometimes result in the same user being payed twice (once by each half), or uncles and aunts being paid by their nephews and nieces in the tree if it has been previously scrambled.  This condition is called \emph{spillover}.

\begin{figure}
    \centering
    \includegraphics[width=0.95\linewidth]{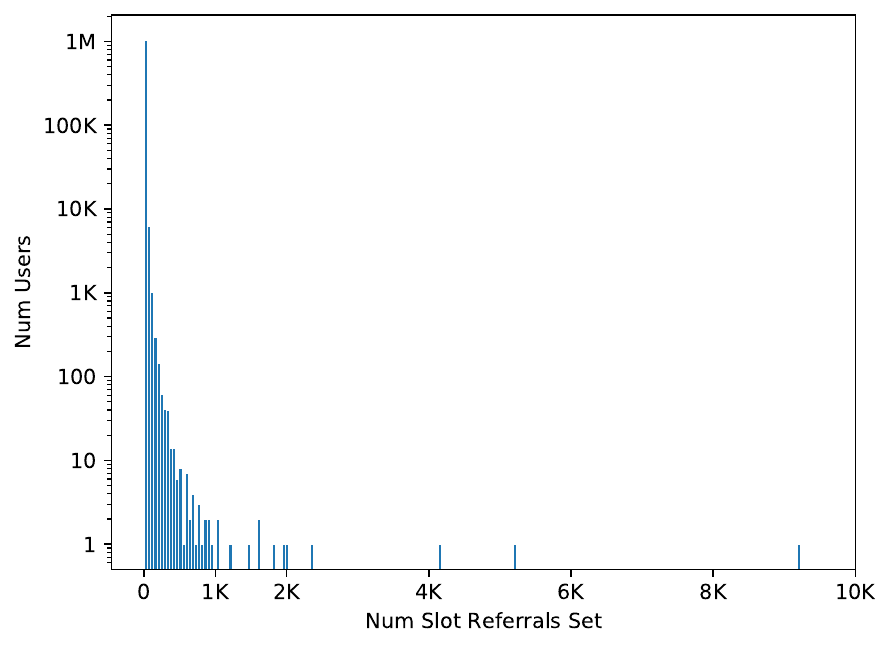}
    \caption{On a log scale, the total number of users (on the y-axis) acting as slot referrer for a given number of \textit{other} users (on the x-axis), for both the X3 and X4 matrices. For example, one user (the contract owner,  \ethaddr{0x81ca1e4de24136ebcf34ca518af87f18fd39d45e}) is slot referrer for 9220 other users.}
    \label{fig:referral_counts}
\end{figure}

Spillover means that it is possible to earn money by receiving payments that should have gone to another user who had blocked slots.
This passive earning is possible only in the X4 system, and only if a spillover recipient's upline is blocked and cannot currently receive payment. 
A given user's chance of spillover is unpredictable, because it depends on the actions of other users. In our analysis of the transactions to Forsage from its conception until January 14th, 2021, we found that 35,251 transactions (only 1.08\%) contained spillover payments. Of those transactions 63\% were registrations, and the remaining 37\% resulted from buying new levels.

Ethereum transaction costs are incurred by each interaction with the Forsage smart contract, and eat into users' profits. Any claims about user profit must thus take gas costs into account.

\paragraph{The privileged role of the owner:}

The Forsage contract is initialized so that the owner account (i.e., the creator of the contract, \ethaddr{81ca1e4de24136ebcf34ca518af87f18fd39d45e
}) has all matrix slots for both X3 and X4 opened for free. Likewise, the owner's slots can never become blocked.  This creates ample opportunities for the owner to profit from the contract, which we confirm empirically in Section~\ref{sec:measurement_study}.

Beyond the ability to earn money by referring users, the owner also has additional opportunities to earn money passively.  If a user sends the contract exactly 0.05 ether for registration without specifically calling the registration function, or calling a function that does not exist, that function call is rerouted to the registration function with the owner set as the user's upline.  Likewise, if the upline gets replaced as the referrer, it is always replaced with a user further up in the pyramid.  Thus, as users refer others and have their slots blocked as a result, the upline for all users eventually converges to the owner of the contract.  Finally, the logic that prevents the owner's slots from becoming blocked also means that the owner's children do not change once set.  This means that the owner maintains the oldest users in the pyramid as children, which results in high spillover in the X4 matrix.

We found that the \textsf{slotReferrer} variable was set to the contract owner for 9220 slots in the Forsage contract. By comparison, the average Forsage user was set as the referrer for 4.14 other accounts (with a standard deviation of 15.92) and the median account was set as the referrer for one other account. Figure~\ref{fig:referral_counts} shows the full distribution of referrers for all accounts.

%% file: Sections/statistics.tex
\section{Contract Measurement Study}
\label{sec:measurement_study}

In this section, we present the results of our measurement study of Forsage contract transactions, which encompasses all monetary transactions in the scheme.  A description of our data collection process is in Section~\ref{para:data_collection}.  We first present statistics capturing the degree of user interaction with the various Forsage contracts on Ethereum and Tron
(\Cref{subsec:stats}). We then present an analysis of the account behaviour and profits over the Forsage user population (\Cref{subsec:most_profitable}), in particular analyzing where funds are obtained and how funds flow through the five most profitable accounts. 

\subsection{Scheme statistics}\label{subsec:stats}

\begin{table*}[h!]
\centering
\small
%\begin{tabular}{ccccc}
\begin{tabular}{m{16mm}m{12mm}m{12mm}m{14mm}m{12mm}m{20mm}m{11mm}}
\toprule
 Contract & Total TXs & Unique sending addresses & Total coins & Total USD & Launch date & Address\\
\midrule
 ETH Matrix  & 3M & 1M & 721k  & 225M & Jan 31, 2020 & \ethaddr{0x5acc84a3e955Bdd76467d3348077d003f00fFB97}  \\ 
  TRX Clone   & 217k &   78k & 537M  & 14M & July 25, 2020 & \trxaddr{TJRv6qukWEz4DKY6gkd3fhX4uahREpTQu6} \\ 
 TRX Matrix  & 1M &  342k &   1B  & 31M & Sept 6, 2020 & \trxaddr{TREbha3Jj6TrpT7e6Z5ukh3NRhyxHsmMug} \\ 
 TRX xGold   & 307k &  105k & 90M  & 2M & Nov 7, 2020 &  \trxaddr{TA6p1BnBf2HJgc77Zk8BHmHoiJzquLCKWb}  \\ 
 ETH xGold   & 37k  &  17k &  8k  & 9M & Jan 4, 2021 & \ethaddr{0x488e3a4bbbb2386ba619eed88319e807c3ddb6c2} \\ 
\bottomrule
\end{tabular}
\caption{Summary statistics of the four official Forsage smart contracts and one clone. The USD value was calculated by taking a sum of the payments per day and multiplying it by the average of the 24-hour high and low on the respective day.}
\label{table:summary_all_contracts}
\end{table*}
% MADE BY THE STATS GANGGGGG

\begin{figure}[t!]
  \centering
    \includegraphics[width=1.0\linewidth]{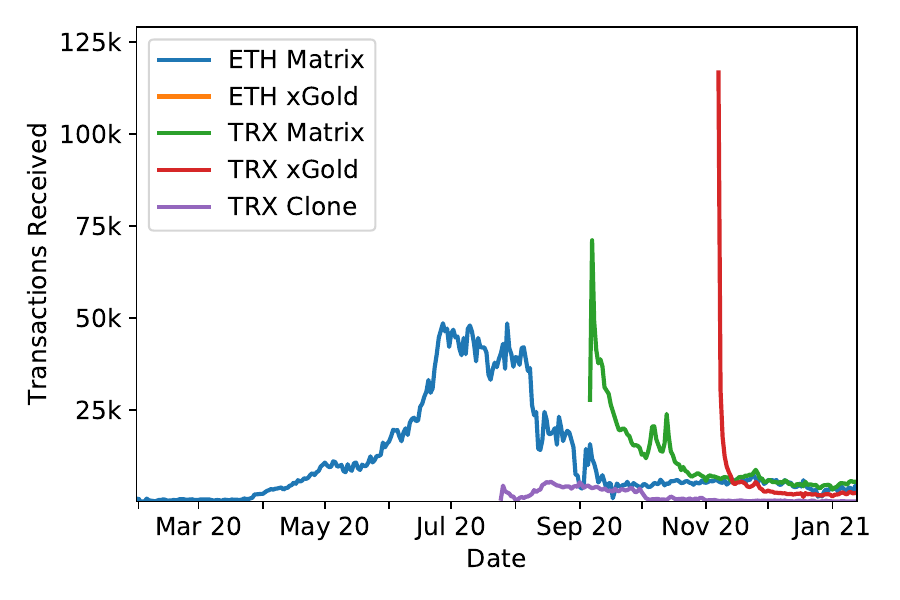}
    \caption{Number of transactions sent from users to the four Forsage contracts across Ethereum and Tron and to an unofficial Tron-based clone.}
    \label{fig:transactionsovertime}
\end{figure}

Table~\ref{table:summary_all_contracts} shows a summary of statistics for the four official Forsage contracts, and one additional contract, TRX Clone, which is a cloned version of the Ethereum Matrix contract operating on Tron.  This clone launched before the official TRX Matrix contract, and has a different domain\footnote{forsagetron.io} but with graphics and style akin to the official website.  The official Forsage website added a warning after the clone's appearance, asking users to ``beware of fake resources'' and stating that the ``forsage.io'' website is the only official domain. 

The table shows that the official Forsage contracts amassed over 267M~USD within the first year of operation. Among all of these contracts, the ETH Matrix contract brought in the most money and raised the highest amount on a single day: 3.7 million USD on August 1, 2020.  
The more recent xGold contracts (deployed on both Ethereum and Tron) were sent a combined 11.53 million USD in ETH and TRX in less than two months.  

Figure~\ref{fig:transactionsovertime} shows the number of transactions received by each contract over time.  For each contract introduced after the original ETH Matrix one, we observe a large number of initial transactions followed by a substantial drop.  We also see a decline in the number of transactions sent to the original ETH Matrix contract after the other contracts become available.  Given the relatively longevity and popularity of the ETH Matrix contract, we focus primarily on it for the remainder of this section.

\begin{figure}
    \centering
    \includegraphics[width=1.0\linewidth]{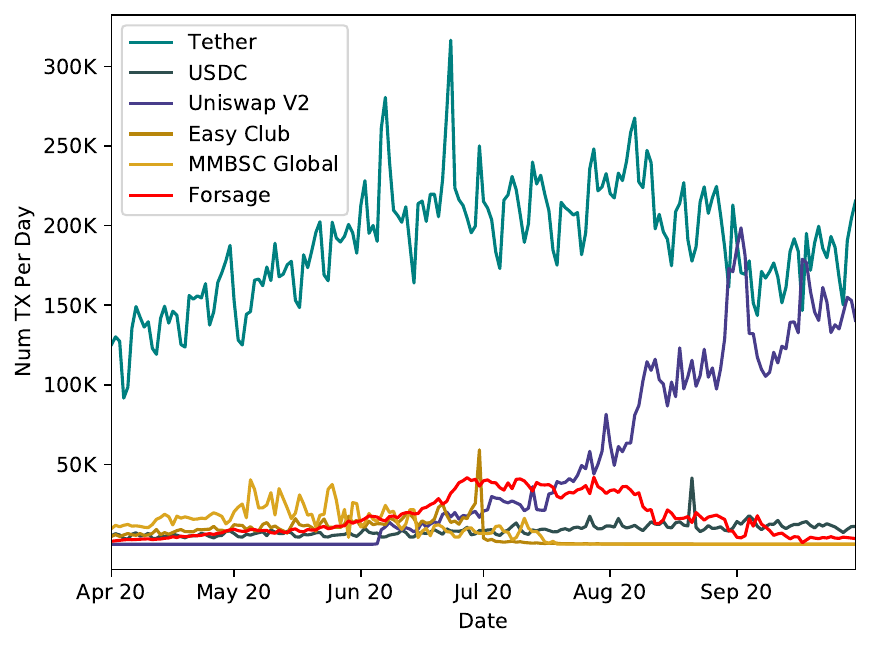}
    \caption{The daily transaction count associated with the six most transacted contracts between April 1 and September 30, 2020.  Here Forsage refers to the ETH Matrix contract.}
    \label{fig:summer_of_scams}
\end{figure}

To illustrate the popularity of Forsage, Figure~\ref{fig:summer_of_scams} shows the number of daily transactions associated with the six most popular contracts across a six-month period in 2020.  Of these contracts, Tether and USDC are stablecoins; Uniswap is a decentralized exchange; and Easy Club, MMBSC Global, and Forsage are believed to be scams/pyramid schemes. We can see that Tether is consistently the most popular contract and that for most of its peak from June to August, Forsage (as represented by ETH Matrix) had the second highest transaction rate among Ethereum smart contracts. This data is supported by Google Trends results for 2020: From April to August of 2020, Forsage had the highest search traffic globally of any of the smart contracts we studied, including both Tether and Uniswap, the two most heavily used smart contracts on the network as of the time of writing.

\iffalse{
\begin{table}
\centering
\begin{tabular}{ccc}
\toprule
Name & ETH & USD \\
\midrule
WoToken.pro & 3.2M & 653M \\
Forsage.io &  721k & 226M  \\
Arbitraging.co & 360k & 52M \\
GoodCycle.io & 64.7k & 15M \\
Million.money & 53.3k & 11M \\
\bottomrule 
\end{tabular}
\caption{Top 5 Ethereum scams with the total Ether and USD received.\tyler{MMMBSC pops up in my chart of most popular smart contracts by number of transactions, why is it not on here / is it the same thing as million.money?}\h{both are pyramid schemes, they are separate, it doesnt appear because it was not in the top 5 in terms of eth received. the MMO scam contract is cloaking its own scam by making its users create new contracts, which then try to hide the money. Is it worth running the numbers for the tokens on that? They are using paxos}}
\label{tab:eth_scam_stats}
\end{table}
}\fi

\subsection{Account behavior and profitability}
\label{subsec:user_behavior}

To understand how Forsage users obtained the funds needed to interact with the contract, we looked at the transactions that sent ETH to their accounts, and at when their accounts first became active.  

\begin{figure}[t!]
	\centering

	\includegraphics[width=1.0\linewidth]{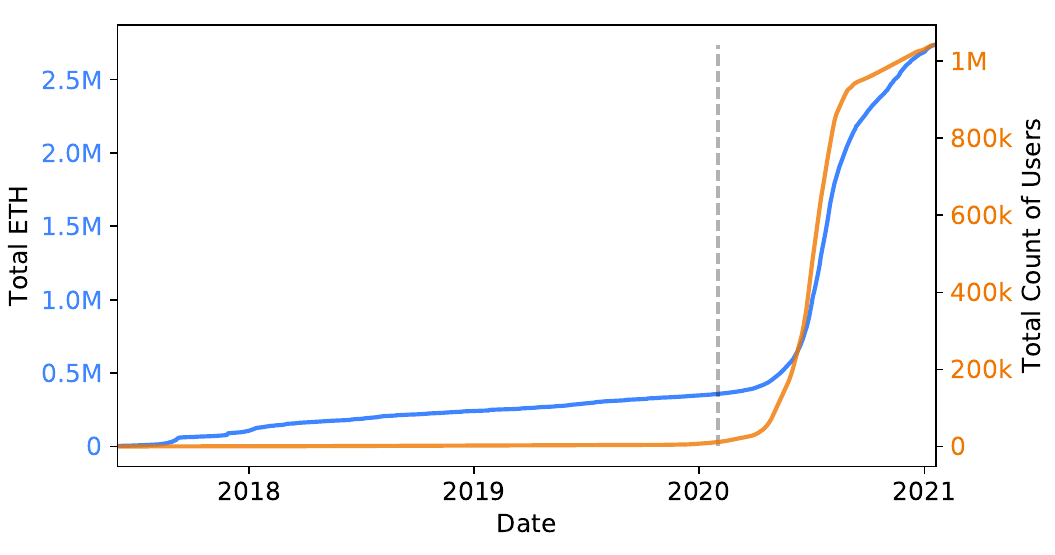}

 	\caption{Total ether received by Forsage users over time and total number of Forsage users according to when their accounts were first used, with a dashed line indicating the Forsage creation date.}
	\label{fig:user_eth_over_time}
\end{figure}

Figure~\ref{fig:user_eth_over_time} shows the ETH received by Forsage users over time and the cumulative count of active Forsage-related accounts (i.e., the first time an account was used that later interacted with the Forsage contract), with a vertical line indicating when Forsage was deployed.  

It is clear that these accounts became active and began to receive substantially more ether after the deployment of Forsage; in fact, 98.89\% of Forsage users had accounts that did not exist (or at least did not transact) before Forsage. 
We found a similar increase when looking at the number of transactions conducted by these users as well: prior to the deployment of Forsage, 11k accounts were involved in 278k 
transactions, but after Forsage's release this increased to 1.04M users engaging in 
16M transactions. 
While the curve in Figure~\ref{fig:user_eth_over_time} looks quite steep given the timescale, it in fact reflects a steady growth in the first appearance of accounts between April and August 2020, which aligns with the peak of Forsage we saw in Figure~\ref{fig:summer_of_scams}.  Each of these months saw thousands of new accounts appearing per day, on average: 1659 in April, 3653 in May, 8272 in June, 10,798 in July, and 4987 in August.  In contrast there were at most 20 new accounts appearing per day for each month in 2019 (except December, when there were 68).

To identify which types of services were the source of this money, we used tags from Etherscan.  Of the ETH sent to Forsage users, over 56\% (1.5M) came from untagged sources, and only 15\% came from known exchanges, with 5\% of this coming from the decentralized exchange Uniswap. As mentioned in Section~\ref{sec:overview}, Forsage promotional material recommends that users obtain ETH from TrustWallet.  This is a non-custodial service, which means accounts are associated with individual users rather than with the exchange.  Thus, if most users followed this advice, we would expect to see that most of the ETH came from untagged sources.

\begin{figure}
    \centering
    \includegraphics[width=\linewidth]{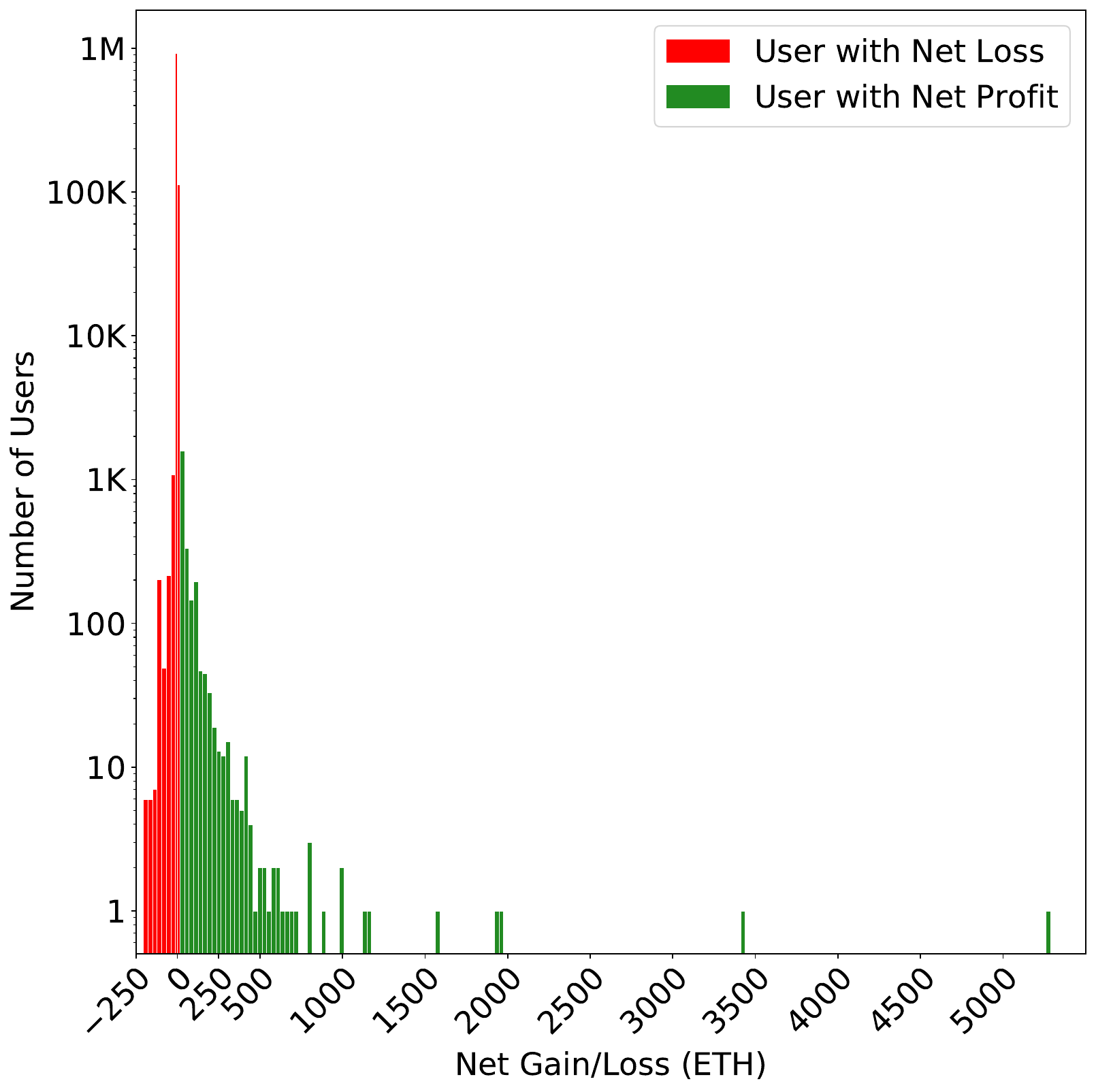}
    \caption{Profit/loss histogram of Ethereum accounts that interacted with the Forsage smart contract, on a log scale. This graph shows the number of accounts that made a profit or loss for each range of ETH. The majority of accounts incurred a small net loss, less than 1 ETH.}
    \label{fig:profit_histogram_all}
\end{figure}

\begin{figure}
    \centering
    \includegraphics[width=\linewidth]{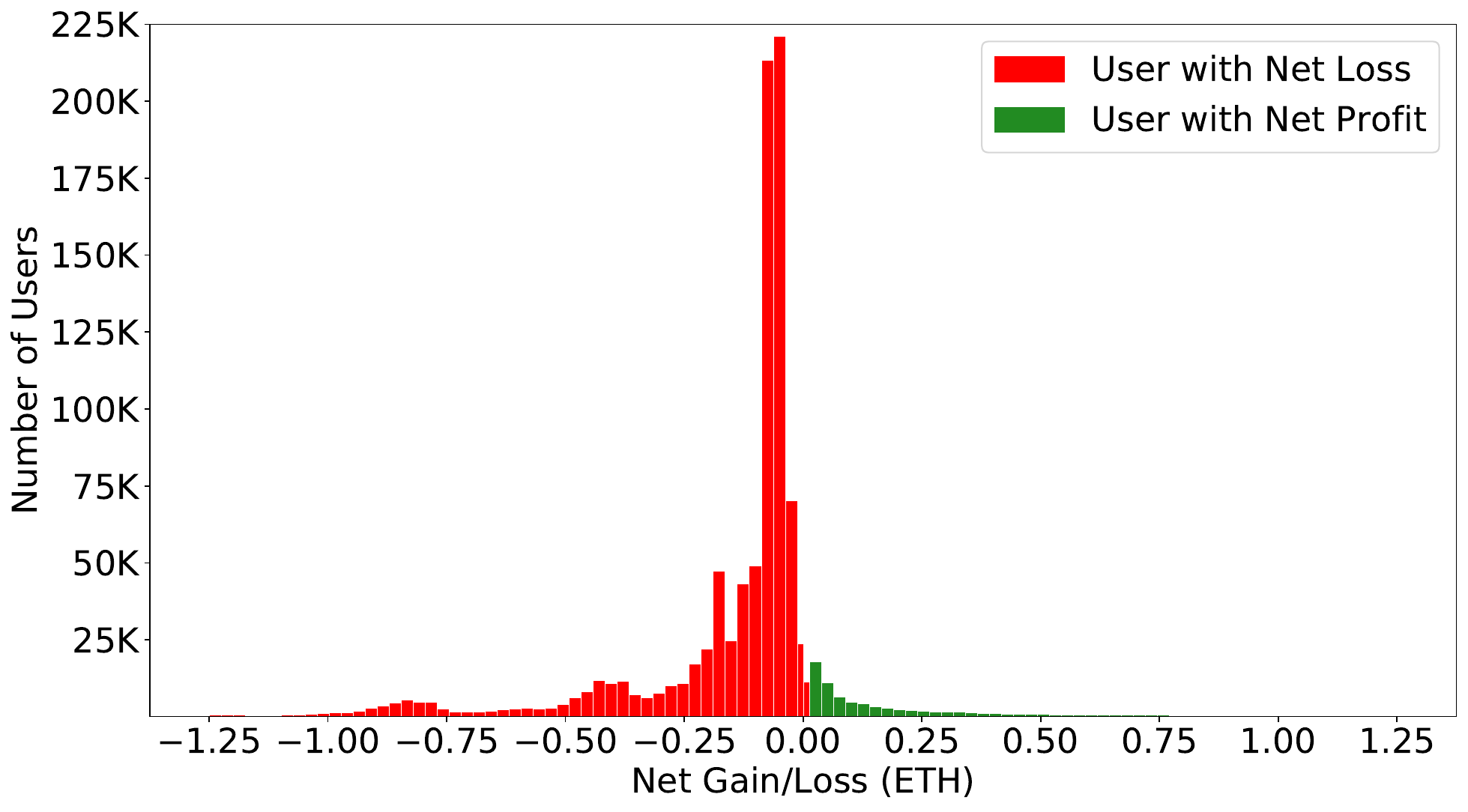}
    \caption{Profit/loss histogram of Ethereum accounts that interacted with the Forsage smart contract, centered around 0 and on a linear scale.  The vast majority of user accounts that interacted with Forsage lost between 0 and 0.25~ETH, with the peak occurring between 0.038 and 0.063~ETH.}
    \label{fig:profit_histogram_zoomed}
\end{figure}

Figures~\ref{fig:profit_histogram_all} and~\ref{fig:profit_histogram_zoomed} show a histogram of all of the accounts that interacted with the ETH Matrix contract organized by the amount of money either gained or lost by each account (including the amount spent on transaction fees) as of January 14, 2021.  In total, of the 1.04 million Ethereum addresses that took part in the ETH Matrix scheme, only 11.8\% (123,979) earned a profit. 
These profitable accounts made 265,618.52~ETH collectively,  %(2.14~ETH on average), 
and the loss-making accounts (919,194 in total) lost 305,785.44~ETH collectively (0.33~ETH on average).  We revisit these profit-making accounts below. Users incur additional losses from the high gas fees paid for transacting with the contract, as explained in Section~\ref{subsec:gas}.

% SERVICES / exchanges 
\iffalse{
\begin{table}
\centering
\begin{tabular}{ccc}
\toprule
Category & ETH & Percentage \\
\midrule
Unknown & 1,553,708.769 & 56.22 \\
High-risk & 717,936.714 & 25.98 \\
Exchange & 275,135.939 & 9.96 \\
Uniswap  & 148,716.116 & 5.38 \\
\bottomrule 
\end{tabular}
\caption{Top 4 categories where forsage users received Ether from. Unknown represents untagged Ethereum addresses. All tags were collected from EtherScan}
\label{tab:eth_tag_stats}
\end{table}
}\fi

\begin{table}
\centering
\begin{tabular}{ccc}
\toprule
 Address & Profit (in ETH) & Notes/First Seen \\
\midrule
 \ethaddr{0x81ca1e4de24136ebcf34ca518af87f18fd39d45e} & 5409.6 & Owner of the contract \\
 \ethaddr{0x44fc2e52243cf20ecc91f61ffa33e59fc7e1c148} & 3445.0 & March 22, 2020 \\
 \ethaddr{0xdedba197cb186e6d129110e71138ef6c6ca153d8} & 1954.9 & March 22, 2020 \\
 \ethaddr{0x4aaa7083535965d1cdd44d1407dcb11eec3f576d} & 1943.2 & January 31, 2020 \\
 \ethaddr{0x59b312f6cfe5b1864654d1942c8c979ad830777e} & 1573.0 & June 4, 2020 \\
\bottomrule
\end{tabular}
\caption{The five most profitable accounts that interacted with Forsage.}
\label{table:most_profitable_accounts}
\end{table}

\paragraph{Profit-making accounts:} 
\label{subsec:most_profitable}

The five addresses with the highest profits in Forsage can be found in Table~\ref{table:most_profitable_accounts}.  Perhaps unsurprisingly given our discussion in Section~\ref{sec:evaluation}, the most profitable Forsage user is the owner of the contract, who earned 5409.6~ETH, or 2.04\% of the total profits.  Collectively, the five most profitable users made 14,325.7~ETH, or 5.4\% of profits, despite representing only 0.0004\% of users.  The top 1000 users made 50\% of the total profits.

Examination of the five most profitable addresses shows that the most profitable address is another Ethereum contract created by the owner of the ETH Matrix contract.  Of the money received by this contract, 99\% came from ETH Matrix.  

The fourth highest earner sent 9\% of received ETH directly back to Forsage.  In fact, if we follow all the addresses to which this user sent money, we see over 1321~ETH sent back to Forsage eventually. 

Similarly, the fifth highest earner sent 204~ETH directly back to Forsage.

Some of the top addresses interact directly with other known scams, such as Beurax.com and TorqueBot.net, meaning they sent or received coins directly from addresses associated with these scams. The top five profit-making accounts received 6.987 ETH from these scams.

Interestingly, the first transaction sent to the address that deployed Forsage was from \ethaddr{0xb19dA4fd9f9A73A5A564C66D229B1E7219e8bdbe}, which is the Ethereum address that deployed Million.money.  This suggests interaction between smart contract-based scam operators.

Finally, we consider the extent to which users who profited by interacting with the Forsage ETH Matrix contract also interacted with other Forsage contracts.  The ETH xGold contract has 17,560 users, of which 17,129 (97.5\%) also interacted with ETH Matrix.  
Furthermore, the highest earner in xGold was the third highest earner in Matrix,  the fourth highest xGold earner was the seventh highest earner in Matrix, and the eighth highest earner in xGold was the second highest earner in Matrix. 
These three earners (all of which are within the ten wealthiest Matrix users) hold 21.85\% of net profits in xGold. 

This suggests that at least some prominent users of Matrix did indeed migrate over to xGold.

%% file: Sections/community.tex
\section{Study of Forsage Community}
\label{sec:community}

\paragraph{Methodology:}
We studied the Forsage community by examining the presence of Forsage on social media. The Forsage website promotes official social media presences on Facebook, Instagram, Telegram, Twitter, and YouTube. All of these services have official APIs to collect data, but some of the research we conducted required manual interaction with the various social websites via a web browser, or more sophisticated data collection techniques like web scraping.

We manually watched YouTube videos to understand the claims that Forsage promotional videos make, as discussed in Section~\ref{sec:youtube}, and made requests to the public YouTube API for view count and other popularity-related data.\footnote{\url{https://developers.google.com/youtube/v3/docs/search/list}} To get a sense of Forsage's Facebook and Instagram presence, we manually browsed various Facebook groups and official Instagram accounts and leveraged the Facebook and Instagram Graph APIs.\footnote{\url{https://developers.facebook.com/docs/graph-api/}} 
Facebook group data is not available on the Graph API so we wrote a custom Python script leveraging the Selenium WebDriver browser automation tool to collect more in-depth data about Forsage Facebook groups and their users.\footnote{\url{https://www.selenium.dev/}} This yielded a dataset of just over 5000 of the most recent members from the largest Facebook group dedicated to Forsage.\footnote{\url{https://www.facebook.com/groups/forsageinformationgroup}}
Using the Twitter API for academic researchers,\footnote{\url{https://developer.twitter.com/en/docs/twitter-api/tweets/search/introduction}} we were able to scrape all tweets with the word ``Forsage" from January 1, 2020 until February 13, 2021. 
We used the official Telegram API \footnote{\url{https://core.telegram.org/}} to collect information about telegram groups related to Forsage.

\paragraph{Community size}
Forsage has a substantial presence on the social network sites that they target. 
This includes:
\begin{itemize}
    \item {\em Facebook:} 131 active Facebook groups with titles or descriptions including {``Forsage,''} containing 403,029 distinct Facebook members.
    \item {\em Instagram:} 24 Instagram accounts with Forsage in the username, disseminating information about Forsage to 24,747 followers of these accounts, with an additional 78,220 posts on the Instagram \#forsage hashtag.
    \item {\em Telegram:} 285,788 people spread across 49 different channels on Telegram dedicated to Forsage.
    \item {\em Twitter:} Our collected Twitter dataset included 85,085 tweets from 21,746 unique accounts, including 513 accounts on Twitter that feature Forsage in the account name. 
    \item {\em YouTube:} 57,551 video results from 325 different YouTube channels. % search query using API forsage ethereum|tron|binance|bsc|cryptocurrency|crypto|matrix space is AND operator pipe is OR
\end{itemize}

The Forsage website also features a ``community'' subdomain\footnote{\url{https://community.forsage.io/}} that hosts a tips and tricks section, blog-post style news, a frequently asked questions section, ``academy courses'' that include video lectures on how to be an effective multi-level-marketer, and a Stack-Overflow-like site where users can ask questions and ``Forsage Community Authors'' answer.

A substantial amount of the Forsage online social media ecosystem may be driven by bots. We ran the University of Indiana's Observatory on Social Media (OSoMe) Botometer tool~\cite{botometertool} on our collected dataset of tweets and found that the tool identified roughly 47\% of the Forsage-related tweets we collected as coming from likely bot accounts. For comparison, in March of 2017, Varol et al.~\cite{botometerstudy} used an earlier version of the Botometer tool to perform a measurement study across all of Twitter and found that ``between 9 and 15\% of active Twitter accounts are bots."

\goodbreak
\begin{table*}[h!]
%\centering
\small
\begin{tabular}{m{15mm}m{100mm}m{20mm}m{20mm}}
\toprule
Type & Claim & \ Appears  & Cumulative Views \\ 
\midrule
Wealth             & Forsage users make money forever.                                               & 3/10    & 425,356 \\
                   & Forsage users make unlimited income.                                            & 3/10    & 449,429 \\
                   & Forsage users make passive income.                                              & 3/10    & 247,344 \\
                   & Forsage users can earn hundreds of ETH in the first few weeks or months.        & 4/10    & 558,617 \\
\midrule
Risk               & Forsage is risk-free for users.                                                 & 3/10    & 393,927 \\
                   & No one can stop Forsage.                                                        & 4/10    & 558,617 \\
                   & Forsage is safe because the contract does not store funds.                      & 4/10    & 530,165 \\
                   & Forsage is scam-proof.                                                          & 3/10    & 393,927 \\
\midrule
Ethereum           & The video explains what Ethereum is for new users.                              & 5/10    & 637,881 \\
Education          & The video explains what a smart contract is for new users.                      & 5/10    & 637,881 \\
\midrule
How to Use         & Successful Forsage users open at least 3 slots per program to start (0.2 ETH).  & 6/10    & 745,960 \\
Forsage            & Users should buy more slots (send Forsage more money) as soon as they earn.     & 5/10    & 654,727 \\
                   & The more slots you open (money you send Forsage), the more you will earn.       & 4/10    & 511,858 \\
                   & If you do not keep opening slots (sending money to Forsage), you will not earn. & 5/10    & 444,539 \\ 
 \bottomrule
\end{tabular}
 \caption{We coded repeated claims that appear across the top 10 most viewed, English language videos on YouTube, which mention "Forsage" in their title to measure user expectations when joining Forsage. }
\label{table:youtubeclaims}
\end{table*}

\subsection{Analysis of Forsage YouTube Promotion}\label{sec:youtube}

Forsage promotional materials offer a window into users' expectations for the contract. They also provides insight into how mention of the technical properties of blockchain technology is harnessed to manipulate novice users. We find that the information gap between those who understand blockchain technology and the broader community provides opportunities for scammers.

\begin{table}
\centering
\begin{tabular}{lccc}
\toprule
Country & Facebook & Twitter & YouTube \\
\midrule
Nigeria & 84 & 4878 & 3 \\
Philippines & 272 & 668 & 14  \\
India & 97 & 488 & 88 \\
United States & 45 & 1019 & 26 \\
Indonesia & 17 & 203 & 8 \\

TOTAL & 771 & 10200 & 216 \\
\bottomrule 
\end{tabular}
\caption{Top five countries with the highest absolute level of Forsage user engagement. User engagement here is measured as a country's total number of Facebook observed users in the most popular Forsage Facebook group, plus its analogous number of Twitter observed users that tweeted about Forsage in 2020, and YouTube data for the number of YouTube channels with geo-tagged locations that produced videos with Forsage in the title of the video.}
\label{tab:socialmedia-absolute}
\end{table}

YouTube is a primary promotional channel for Forsage. Each participant joining Forsage is referred to an official YouTube video explaining the program~\cite{youtube1}. We searched YouTube for English language videos with ``Forsage'' in the title and tracked the claims that repeat across videos to measure user expectations for Forsage. The search for most viewed videos about Forsage also returned promotional videos in Tagalog, Russian, Hindi, Tamil, Bangala, Telugu, Indonesian, and Spanish. Quasi-official (they share the same branding) Telegram chat groups for Forsage news exist in English, Spanish, French, Italian, Russian, Arabic, Portuguese, Hindi, Tamil, German, Azerbaijani, and Turkish. 

Recommendation algorithms, like the one used by YouTube for search results, work in terms of popularity measured in views. The most viewed videos on YouTube are the most likely to be seen by users. We selected the top ten videos by views to qualitatively measure what users who search for informational videos about Forsage would see and hear about the program and gain a sense of participant expectations. We did so by coding the claims asserted about Forsage in these videos. We focused on just the top ten videos because coding claims is a labor-intensive, manual process. A researcher watched each video and noted if each video contained any instance of certain claims (see Table~\ref{table:youtubeclaims} and Appendix~\ref{app:youtube}). Each video was watched and coded twice to ensure accuracy.

The top ten YouTube videos we coded had between 267,008 views (1st) and 61,996 views (10th). Beyond the videos we coded, the 11th most viewed video had just over 50,000 views\footnote{\url{https://www.youtube.com/watch?v=aGi5G5mTCUM}} 
and the 20th had 33,000 views.\footnote{\url{https://www.youtube.com/watch?v=9vlOYRSLaHI}}

The top 10 ``Forsage'' YouTube videos by views as of December 14, 2020 (see Appendix~\ref{app:youtube}) fit into three categories: official promotion, user-led recruitment, and user reviews. Two of the videos were official promotion posted to Forsage's YouTube channel~\cite{youtube1, youtube6}. Table~\ref{table:youtubeclaims} shows the repeated claims across the top ten videos. 

In recruiting new users, Forsage promoters pointed to users who earned tens of thousands of dollars per day and hundreds of thousands of dollars per month, showing images of successful users' Forsage dashboards displaying six-figure returns. %\cite{youtube5, youtube6}. 
Forsage official promotion videos highlight the immutable nature of the smart contract and the transparency of Ethereum as proof that Forsage cannot be a scam.  %\cite{youtube1, youtube6}. 
They also make claims about the life-changing wealth and unstoppable, passive income that users could unlock from the Forsage contract. %\cite{youtube6}. 

Forsage promotional videos also provide basic explanations of blockchains, Ethereum, smart contracts, and how to use a cryptocurrency wallet to pay the contract, implying that they expect users to be cryptocurrency novices. %\cite{youtube1, youtube2}.
Only one of the top ten videos identifies Forsage as a scam and warns users against using it.  % \cite{youtube8}.

Many of the incorrect claims made in the Forsage promotional YouTube videos also appear on the Forsage website and in the questions section of the official Forsage Community website. 
 
\goodbreak
\begin{figure*}[t!]
  \centering
    \includegraphics[width=0.99\linewidth]{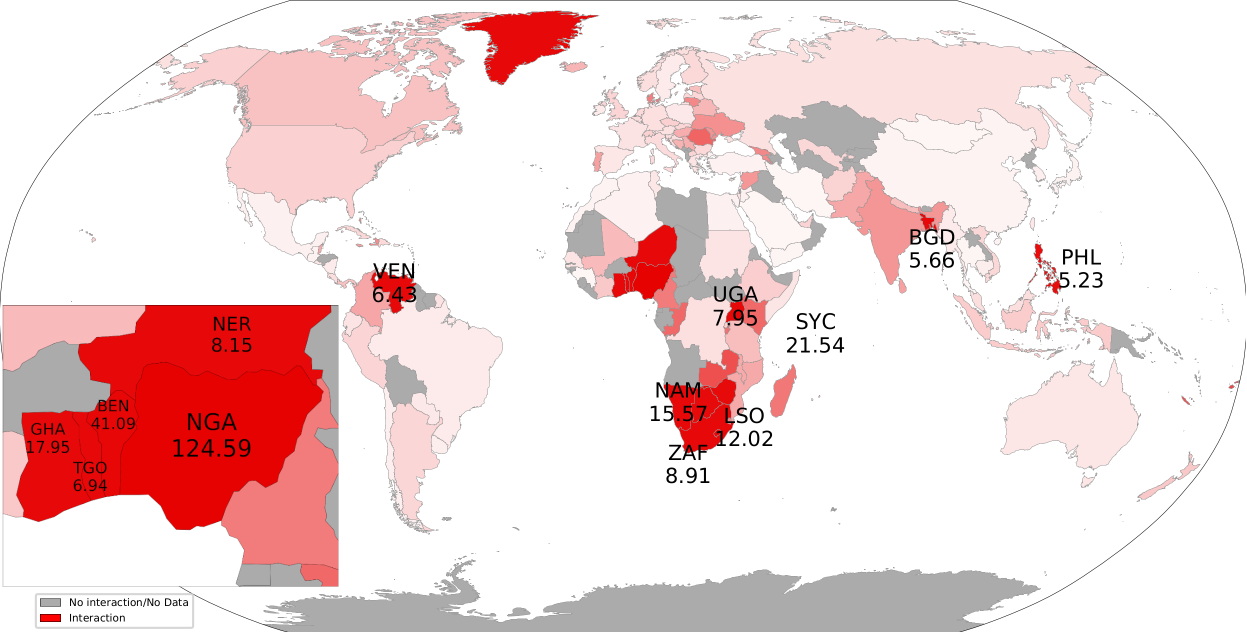}
    \caption{Forsage social media interaction heat map by country. Country labels indicate the ISO-alpha-3 name of the country and the number of Forsage users per 100k people in that country.  The data reflects the public location of members in a popular Forsage Facebook group and Twitter users that tweeted about Forsage. Countries depicted in gray had no Forsage interaction. The intensity of color from white to red is scaled linearly from the 0th percentile of data to the 90th percentile, and everything above 90\% of the data is colored the same shade of dark red. This slightly understates the relative depth of penetration in outlier countries like Nigeria. For the full list of countries and their scores, see Appendix \ref{sec:socialmediascoreappendix}.}
    \label{fig:heatmapsocialmedia}
\end{figure*}

\subsection{Forsage user geography}
\label{sec:community-geography}

Since transactions on the Ethereum network do not carry any inherent geographic metadata, we turned to social media analysis in order to gain a sense of the geographic placement of people interested in Forsage.  In the data we collected on members of Forsage-related Facebook groups, we found 771 users that publicly listed a country location on their Facebook profile. We also found 10,200 unique Twitter accounts that publicly posted their geographic location.  YouTube does not expose information about geographic location of the consumers of YouTube videos, but YouTube channels that produce videos can choose to include country location in their channel profile.  We summarize this data for the five countries with the highest number of active users in Table~\ref{tab:socialmedia-absolute}.  Despite having a substantial population and being the  nationality of the founders of Forsage, Russia was not a large source of Twitter or Facebook content, although the country did produce a large number of YouTube videos and content about Forsage.

The high number of Forsage users in the Philippines may explain why the Philippines SEC took action to raise awareness about the malicious intent behind Forsage~\cite{psec_warning_2020,psec_forsage_2020}, unlike other countries. Likewise, Nigeria has high penetration rates for both cryptocurrency and Forsage, and has recently banned cryptocurrency payments from its banking sector~\cite{nigeriabanscrypto}.
While each of these five countries had high Forsage activity in absolute terms, they also have large populations.  
We thus normalized our Facebook and Twitter data relative to the specific populations on each service for each country (i.e., the number of people per country divided by a public estimate of the number of Facebook and Twitter users in that country) to get a sense of the number of Facebook and Twitter users, per 100,000 users, that interacted on each platform with the Forsage topic. Statistics for the number of Facebook and Twitter users per country came from Miniwatts Marketing Group, WeAreSocial, and Hootsuite~\cite{miniwatts, hootsuite}. We did not include the YouTube data at this stage as it was too small to be useful.  We gave equal weight to the numbers for Facebook and Twitter to produce the heat map in Figure~\ref{fig:heatmapsocialmedia}.

Our normalized data showed that Forsage is most popular in Nigeria and the African continent, the Philippines, and Venezuela. Greenland, the Seychelles, and some Caribbean islands may appear to have heavy Forsage penetration, but may be outliers due to small population sizes. Google Trends traffic and geographic data agree with our conclusions: Google Trends shows the greatest amount of population-adjusted search traffic in Nigeria and surrounding West African countries, and shows a peak in user search interest in July 2020, which is when we observed a similar peak in transactions involved Forsage in Figure~\ref{fig:transactionsovertime}.

Familiarity with cryptocurrency does not appear to have any positive or negative correlation with interest in Forsage: The 2021 Statista Global Consumer Survey~\cite{cryptopercountry} lists the top countries globally with the reported highest number of cryptocurrency users. 
Vietnam (\#2) and China (\#3) both had relatively high levels of cryptocurrency use, but low levels of interest in Forsage.   Similarly, familiarity with cryptocurrency does not appear to prevent people from falling for the Forsage scam, as in the case of Nigeria and the Philippines (\#1 and \#3 globally for cryptocurrency usage). Nigeria may be a special case, as Statista found that almost a third of Nigerians said they used cryptocurrency, far beyond most countries. It is also an outlier in the data for interest in Forsage. 

%% file: Sections/howToFix.tex
\section{Proposed Solutions}
\label{sec:solutions}

\subsection{Targeted education}
From our analysis of Forsage user locations in Section~\ref{sec:community-geography}, the majority of Forsage victims are located in only a few countries. This concentration lends itself well to a targeted education campaign and warnings from local financial leaders about the Forsage scam.  For example, a simple user dashboard showing the number of Forsage users who lose money from the contract---more than 88\% as of January 15th, 2020---could serve as an effective tool to combat disinformation from Forsage promoters about the wealth users can amass. Such statistics may be more effective than general warnings such as that issued by the Philippines SEC (see below).

\subsection{Law enforcement and regulation}

Past cryptocurrency pyramid schemes, including Plustoken, Wetoken, Onecoin, and Bitconnect, have collapsed as a result of government sanctioning, which included the arrest or warrants for the arrest of the founders and leadership \cite{palmer_2020, haig_2020, bel_2020, madeira_2020}.  Similar attempts have been made around the world in regards to Forsage.  On June 30, 2020, The Philippines Securities and Exchange Commission (PSEC) issued numerous warnings declaring that Forsage was not a registered entity within their jurisdiction and was operating without a license. 
On September 30, 2020 the PSEC released a public announcement, mentioning that Forsage was publicly selling securities as investment contracts without a license \cite{psec_cadpublic_2020, psec_cadpublic_state_2020, psec_forsage_2020, psec_warning_2020}. The PSEC served a cease-and-desist order. Forsage refused to comply, responding that they
``are outside the Commission’s jurisdiction.'' 
On March 22nd 2021, the Commissioner of Securities and Insurance of the U.S. state of Montana ordered Forsage to cease and desist from operating a pyramid scheme in Montana~\cite{montanaDA, montanaDA2}. 

To date, the authors of this paper are unaware of any public arrests made in relation to the Forsage contract. The contract authors continue to profit and their Ethereum addresses actively submit transactions to the network.

\subsection{Voluntary blocklisting}

Previous research has shown blocklisting can effectively combat scams and illicit activity. Moser et al. found that transaction blocklisting of illicit cryptocurrency funds is an effective additional layer above existing anti-money laundering (AML) and know-your-customer (KYC) requirements for cryptocurrencies~\cite{moser2019effective}. Previous research in illicit online pharmaceutical sales found that the payment processing services are the most fragile part of the scam~\cite{180200}. These services play a similar role in online pharmaceutical sales to fiat-accepting cryptocurrency exchanges in Forsage, suggesting that access to exchanges, which could be revoked with blocklisting, may be the the most fragile part of the scam.  

Crypto Defenders Alliance (CDA)\footnote{\url{https://cryptodefendersalliance.com/}} and CryptoSafe Alliance\footnote{\url{https://www.cryptosafe.org/}} are two examples of groups that operate a blocklist. 

On the other hand, blocklists can be biased and enable forms of censorship, and addresses that are blocked in one region may not be considered suspicious or criminal in other regions.  
To understand how professionals navigate these tensions, we spoke to an anti-money laundering cryptocurrency investigator at a high profile exchange.  This expert expressed a belief that it is the responsibility of law enforcement and regulators to comment 
on whether or not an address should be blocked, and that it would be unfair and unjust to hold a user's funds without an explicit request from law enforcement or a court of competent jurisdiction.  Nevertheless, some exchanges have joined the alliances mentioned above, due to the time and resources required to maintain a dedicated list of blocked addresses themselves.

%% file: Sections/relatedWork.tex
\section{Related Work}

Past research has sought to quantify and characterize scams running on Ethereum and in Bitcoin. For Ethereum based scams, Chen et al. used data mining and machine learning to detect Ponzi schemes ~\cite{chen_zheng_cui_ngai_zheng_zhou_2018} while Yu et al. modeled Ponzi scheme identification and detection as a node classification task ~\cite{yu2021ponzi}. Bartoletti et al. characterized and summarized the Ethereum Ponzi scheme ecosystem by comparing the code and promotion for Ethereum smart contract schemes ~\cite{BARTOLETTI2020259}. They found that scammers use the guaranteed execution of smart contracts public nature of  Ethereum to inspire confidence in their victims.  
For Bitcoin, Vasek et al.~\cite{Vasek2015TheresNF} and Bartoletti et al.~\cite{Bartoletti2018} both worked to detect and model Bitcoin-based scams. These included Ponzi schemes that collect Bitcoin from victims, where Vasek et al. found that more than half of scams last less than one week. 
Paquet-Clouston et al.~\cite{Paquet2019} and Xia et al.~\cite{xia2020dont} studied specialized scams that leverage Bitcoin for payments: respectively, scams to extort payments using threats of revealing intimate data, and scams claiming to raise money for COVID-19 research and relief. 

Other research has identified existing types of scams that have now migrated onto blockchains.  One example is pump-and-dump schemes, 
where scammers use online chat services like Discord and Telegram to manipulate the price of cryptocurrencies and then sell their holdings of those cryptocurrencies for profit~\cite{236350, Hamrick2018, Kamps2018}. Other scams are new to blockchains and exploit their unique characteristics.
One example is honeypot scams. These are executed using honeypot smart contracts, which are built to include financial traps within the contract itself~\cite{Ferreira2019}. 

In past work characterizing the victims of blockchain-enabled scams, Phillips et al.~\cite{phillips_wilder_2020} showed that victims tend to send funds from fiat-accepting cryptocurrency exchanges, making the scams accessible to novice cryptocurrency users. They also found that scammers often create multiple, similar scams, which run in parallel. Yousaf et al. showed that scammers use shifting services to convert Ether into other coins in an attempt to thwart tracking~\cite{236358}.

%% file: Sections/conclusion.tex
\section{Conclusion}

We presented an in-depth measurement study of Forsage, a smart-contract pyramid scheme. Forsage is currently active and was at one time the second most actively used contract in Ethereum. 

We found that community claims regarding the open and verifiable nature of Forsage are belied by the contract's considerable complexity.
Our study consequently required a number of different data gathering approaches. It also required the creation of new tools---of potential independent interest and to be open-sourced---to analyze the state of the Forsage contract. Thanks to these tools, our study provides detailed insights into the mechanism design, transaction costs, and other features of Forsage.  

Among our key findings were that the vast majority of Forsage accounts---over 88\%---incurred losses, for a combined total loss of 305,785 ETH. The contract owner, in contrast, earned over 5000 ETH (well over 1M~USD), while a small number of other accounts at the top of the pyramid earned similarly large sums. 

Our analysis of Forsage promotional materials reveals that scammers in the Forsage community have taken advantage of misconceptions and misinformation about blockchain technology, using properties like open-source code and transaction transparency as a source of legitimacy with users who lack the skills necessary to understand the  contract's behavior. Our analysis of Forsage on social media shows geographically distinct communities of scammers and victims, with the scammers based primarily in Russia and victims apparently located mainly in Nigeria, southern Africa, the Philippines, Venezuela, Indonesia, and India.

Public warnings about Forsage by entities such as the Philippines SEC have had little apparent effect. We show that Forsage creators have launched new and currently lucrative Forsage variants, some now on blockchains other than Ethereum. We hope that our findings can help stem this spread. In addition to providing insights that may serve to educate potential victims, our study demonstrates highly concentrated earnings among top-earning accounts, suggesting that targeted blocklisting could be an effective step to slow the growth of Forsage and contracts like it. 

%% file: Sections/acknowledgments.tex
\goodbreak

\section{Acknowledgments}

We would like to thank Chainalysis for use of their Reactor tool in support of this research.

This work was funded by NSF grants CNS-1564102, CNS1704615, and CNS-1933655 as well as generous support from IC3 industry partners. Any opinions, findings, conclusions, or recommendations expressed here are those of the authors and may not reflect those of these sponsors.

%% file: Sections/appendixA.tex
\clearpage

\section{Social Media Scores}
\label{sec:socialmediascoreappendix}

Note that in the following table, any country that is not listed had no observed interactions with Forsage in our data set discussed in section \ref{sec:community}.

\tablefirsthead{\toprule Country Code&\multicolumn{1}{c}{Forsage Interactions per 100K People} \\ \midrule}
\tablehead{%
\multicolumn{2}{c}%
{{\bfseries Continued from previous column}} \\
\toprule
Country Code&\multicolumn{1}{c}{Forsage Interactions per 100K People}\\ \midrule}
\tabletail{%
\midrule \multicolumn{2}{r}{{Continued on next column}} \\ \midrule}
\tablelasttail{%
\\\midrule
\multicolumn{2}{r}{{Concluded}} \\ \bottomrule}
\begin{supertabular}{ll}
		country & score \\
		NGA & 124.594 \\
		BEN & 41.087 \\
		SYC & 21.538 \\
		GHA & 17.953 \\
		NAM & 15.573 \\
		LSO & 12.017 \\
		WSM & 11.364 \\
		GRL & 8.929 \\
		ZAF & 8.910 \\
		NER & 8.152 \\
		UGA & 7.952 \\
		BMU & 7.042 \\
		TGO & 6.944 \\
		VEN & 6.426 \\
		BGD & 5.657 \\
		ZWE & 5.492 \\
		BWA & 5.483 \\
		PHL & 5.227 \\
		VCT & 4.310 \\
		FJI & 3.906 \\
		SXM & 3.788 \\
		ZMB & 3.747 \\
		NCL & 3.676 \\
		ROU & 3.287 \\
		COG & 3.193 \\
		KEN & 3.153 \\
		BDI & 2.976 \\
		MDG & 2.841 \\
		CMR & 2.764 \\
		DNK & 2.670 \\
		ATG & 2.551 \\
		MLT & 2.473 \\
		LTU & 2.469 \\
		GEO & 2.469 \\
		BHS & 2.453 \\
		SWZ & 2.451 \\
		ARM & 2.381 \\
		HTI & 2.347 \\
		UKR & 2.300 \\
		MDA & 2.232 \\
		IND & 2.150 \\
		SYR & 2.128 \\
		JAM & 1.964 \\
		PRT & 1.933 \\
		LKA & 1.890 \\
		TTO & 1.805 \\
		COL & 1.804 \\
		PAK & 1.786 \\
		MOZ & 1.736 \\
		MWI & 1.712 \\
		BLZ & 1.689 \\
		HUN & 1.645 \\
		SRB & 1.614 \\
		CYM & 1.562 \\
		BIH & 1.535 \\
		ISL & 1.475 \\
		BRB & 1.445 \\
		BLR & 1.429 \\
		MUS & 1.344 \\
		MLI & 1.344 \\
		TZA & 1.262 \\
		NPL & 1.261 \\
		EST & 1.221 \\
		PAN & 1.173 \\
		CAN & 1.166 \\
		MDV & 1.108 \\
		NZL & 1.040 \\
		CIV & 1.038 \\
		IDN & 0.959 \\
		ETH & 0.902 \\
		HKG & 0.876 \\
		RWA & 0.874 \\
		USA & 0.867 \\
		CZE & 0.863 \\
		CHE & 0.833 \\
		AFG & 0.784 \\
		ARE & 0.777 \\
		MKD & 0.775 \\
		GBR & 0.736 \\
		PER & 0.736 \\
		ARG & 0.729 \\
		SVK & 0.718 \\
		SGP & 0.702 \\
		LVA & 0.697 \\
		CUB & 0.690 \\
		VNM & 0.672 \\
		UZB & 0.672 \\
		NLD & 0.659 \\
		BGR & 0.655 \\
		DOM & 0.650 \\
		FIN & 0.647 \\
		ITA & 0.604 \\
		HRV & 0.591 \\
		BEL & 0.580 \\
		SVN & 0.573 \\
		CYP & 0.552 \\
		NOR & 0.541 \\
		MYS & 0.531 \\
		CRI & 0.527 \\
		COD & 0.505 \\
		DEU & 0.492 \\
		AUT & 0.485 \\
		RUS & 0.470 \\
		ISR & 0.462 \\
		FRA & 0.439 \\
		ECU & 0.405 \\
		SDN & 0.403 \\
		SEN & 0.401 \\
		GRC & 0.401 \\
		SOM & 0.388 \\
		ESP & 0.376 \\
		TUN & 0.374 \\
		AUS & 0.368 \\
		SLV & 0.336 \\
		POL & 0.324 \\
		PRY & 0.312 \\
		TWN & 0.311 \\
		MMR & 0.284 \\
		NIC & 0.276 \\
		SWE & 0.270 \\
		IRL & 0.267 \\
		BRA & 0.244 \\
		PRI & 0.216 \\
		KHM & 0.184 \\
		URY & 0.182 \\
		KOR & 0.171 \\
		CHL & 0.162 \\
		QAT & 0.154 \\
		EGY & 0.138 \\
		JOR & 0.134 \\
		MEX & 0.133 \\
		TUR & 0.127 \\
		MAC & 0.123 \\
		DZA & 0.118 \\
		GTM & 0.101 \\
		JPN & 0.096 \\
		LBN & 0.085 \\
		KWT & 0.083 \\
		CHN & 0.075 \\
		MAR & 0.075 \\
		BHR & 0.074 \\
		THA & 0.061 \\
		YEM & 0.050 \\
		IRN & 0.027 \\
		GIN & 0.025 \\
		MNG & 0.022 \\
		SAU & 0.017 \\
\end{supertabular}

%% file: Sections/appendixB.tex
\clearpage

\onecolumn{
\section{Top YouTube Videos}\label{app:youtube}

% \usepackage{booktabs}
{\small
\begin{table}[hb]
\begin{tabular}{S[table-format=2.0]lS[table-format=6.0]l}
\toprule
Rank & Title                                                                 & {Views}   & Link                                                \\ \midrule
1    & Forsage Overview: Earn Ethereum Daily!                                & 267008 & https://www.youtube.com/watch?v=m0NzYwFfGH4         \\
2    & Forsage Presentation - How does Forsage work                          & 120425 & https://www.youtube.com/watch?v=NoAh57M-Dak         \\
3    & Forsage Smart Contract- \$735 Made Without Referring Anyone           & 113677 & https://www.youtube.com/watch?v=PqsfdcLvlIQ         \\
4    & FORSAGE: HOW TO EARN WITHOUT RECRUITING ANYONE IN FORSAGE             & 117931 & https://www.youtube.com/watch?v=zCvj\_zZZmOI        \\
5    & Forsage Smart Contract Review - Is It A SCAM Or Legit Ethereum MLM?   & 106261 & https://www.youtube.com/watch?v=7YUWfl0looY \\
6    & FORSAGE.io - BIG SPECIAL EVENT                                        & 91973  & https://www.youtube.com/watch?v=NMfcDSCXLK8         \\
7    & Forsage Smart Contract \$1,778 Made Without Referring a Single Person & 91188  & https://www.youtube.com/watch?v=N-Qem777Qis         \\
8    & Forsage Review - Is Forsage a Scam or Legit?                          & 79264  & https://www.youtube.com/watch?v=WVsLIVCqJbc         \\
9    & Smartway Forsage REVIEW - First Ever SCAM PROOF Program               & 64923  & https://www.youtube.com/watch?v=TmVp2ViU0Ro         \\
10   & how to make money on forsage without referring anyone                 & 61996  & https://www.youtube.com/watch?v=qTVMCjuipho         \\ \bottomrule
\end{tabular}
\end{table}
}
}

%% file: Sections/AppendixC.tex
%\clearpage

%\onecolumn{
\section{Payment Flow Chart for X4}

\begin{figure*}[hb]
\centering
\includegraphics[width=0.95\textwidth]{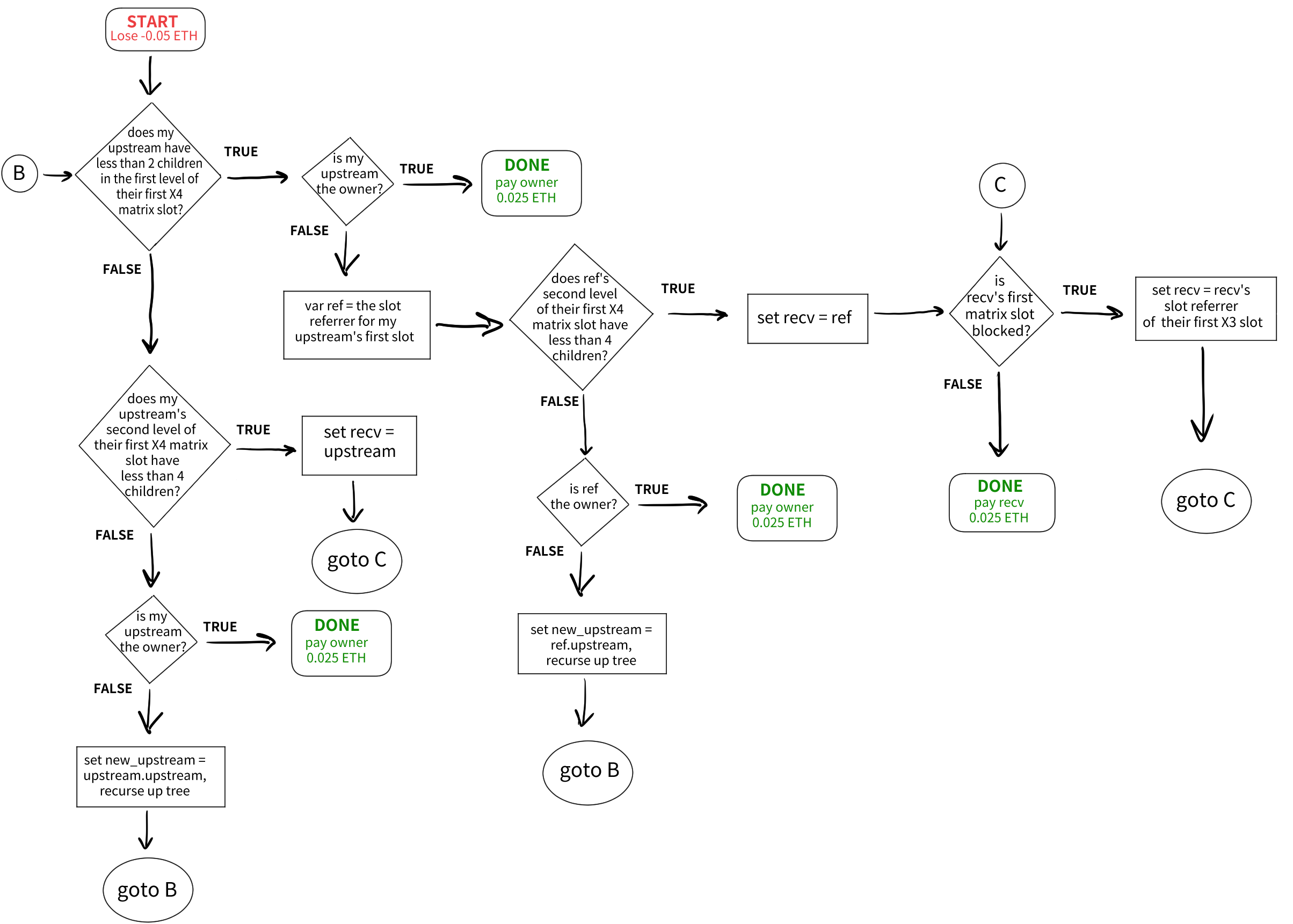}
\caption{Flow chart for the logic of who gets paid when a new user registers, in the X4 system. The \texttt{BuyNewLevel} function follows similar logic, but would operate on the matrix slot being purchased for conditionals, rather than the first slot.}
\label{forsage-payflow-x4}
\end{figure*}
%}

%% file: Sections/AppendixD.tex
\section{Forsage Website Screenshots}
\label{sec:webpagescreenshots}

\newcommand{\subf}[2]{%
  {\small\begin{tabular}[t]{@{}c@{}}
  #1\\#2
  \end{tabular}}%
}

\begin{figure}
\centering
\begin{tabular}{cc}
\subf{\includegraphics[width=85mm]{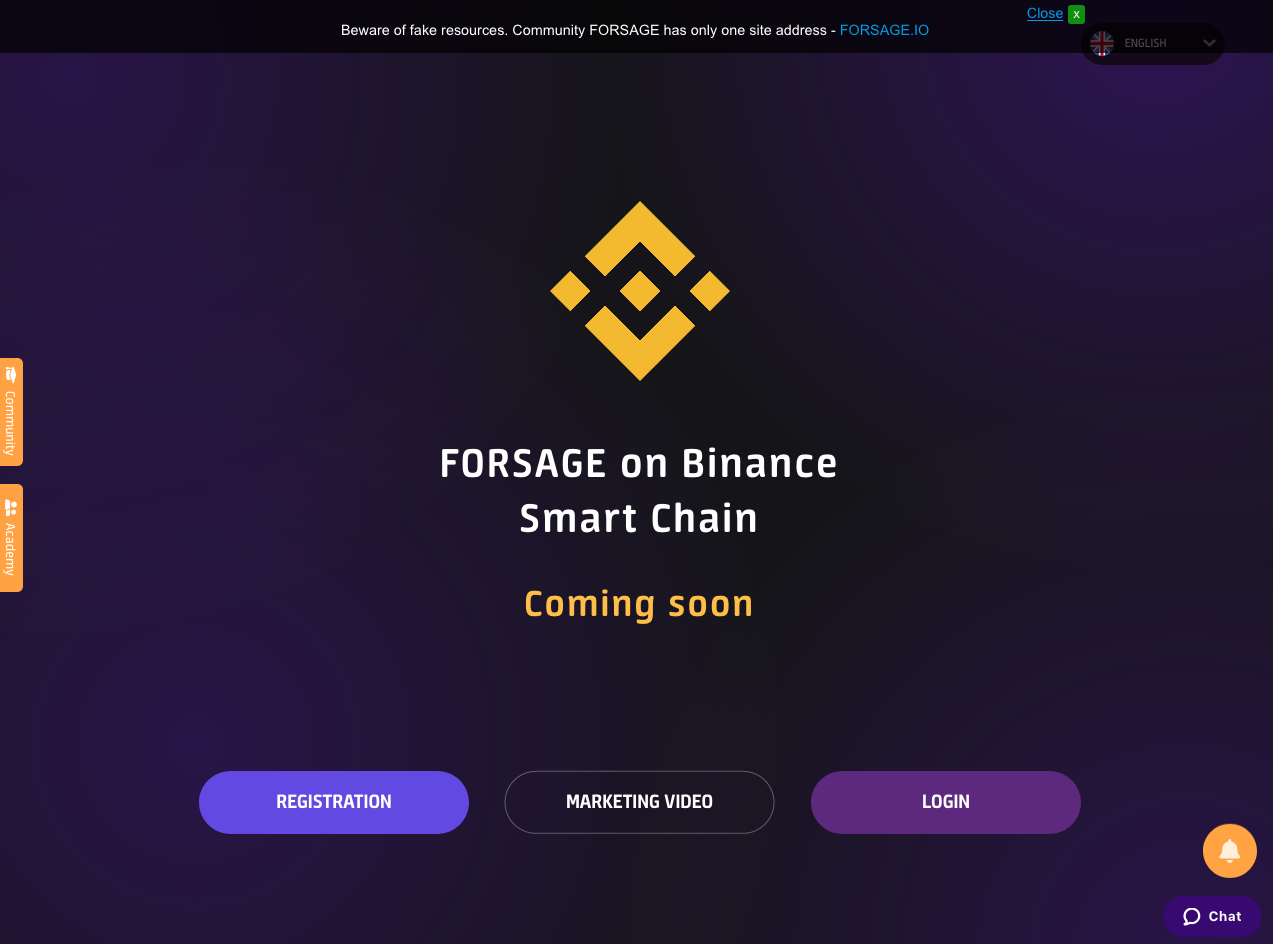}}
     {(a)}
&
\subf{\includegraphics[width=85mm]{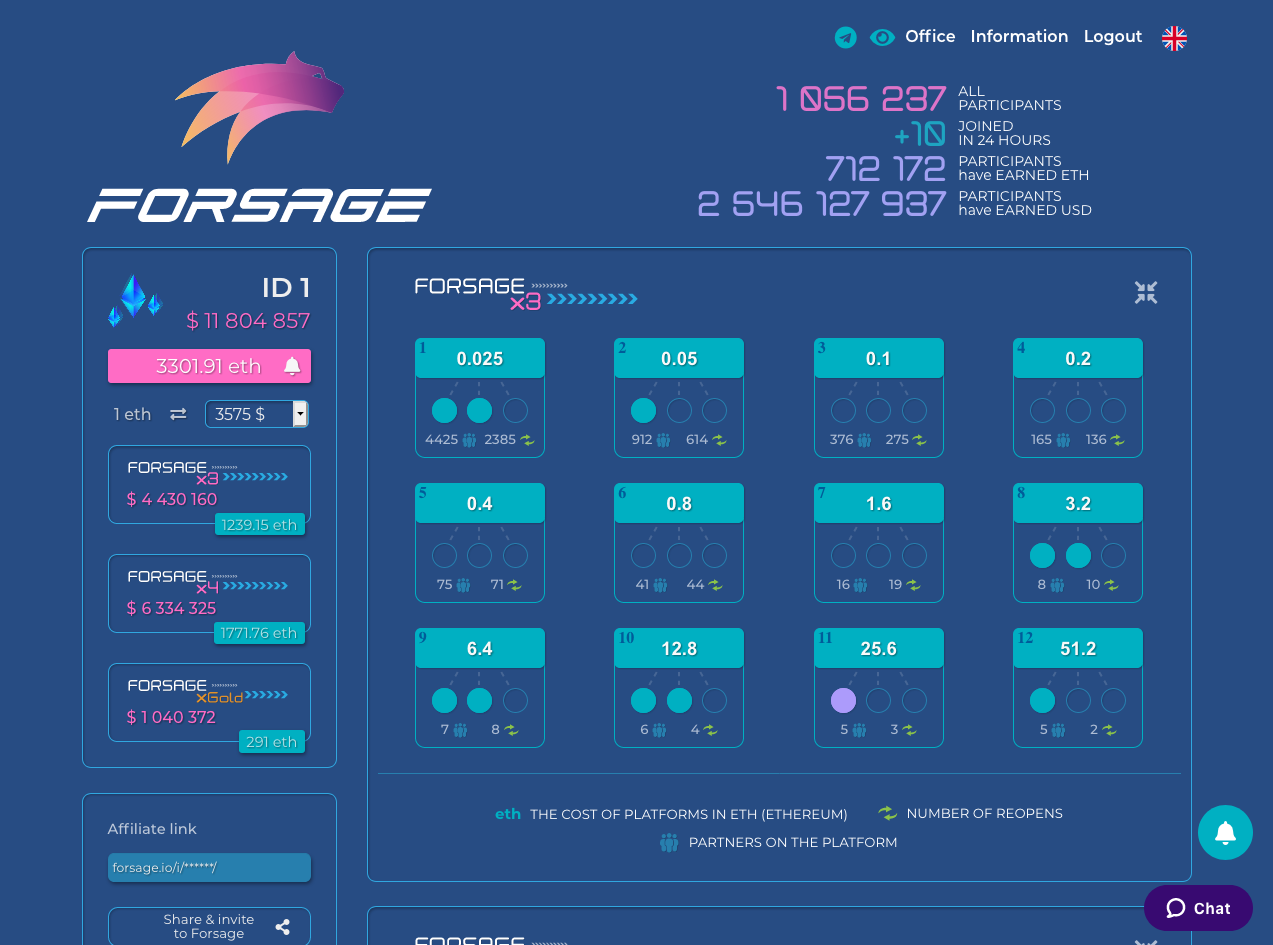}}
     {(b)}
\\
\subf{\includegraphics[width=85mm]{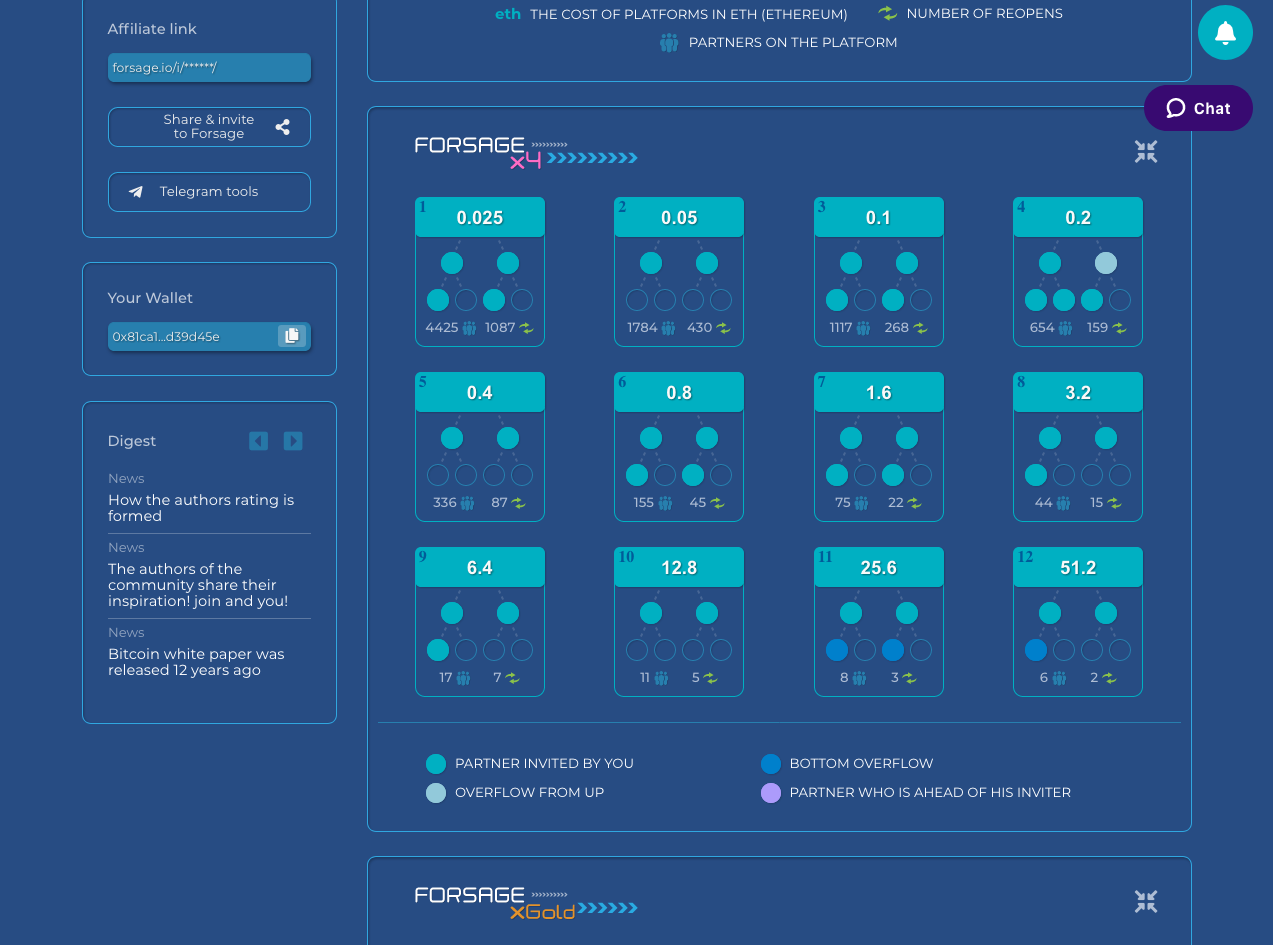}}
     {(c)}
&
\subf{\includegraphics[width=85mm]{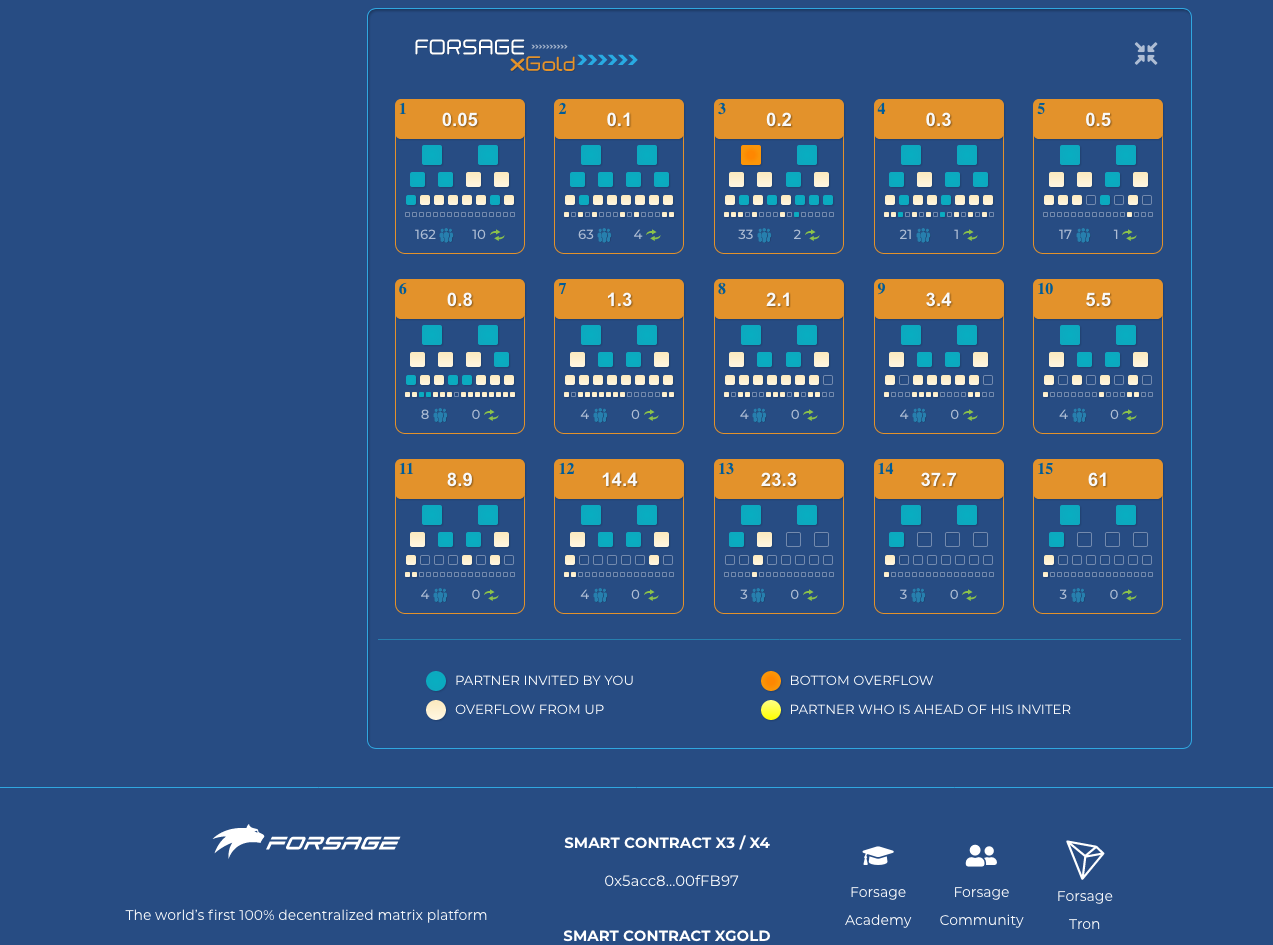}}
     {(d)}
\\
\end{tabular}
\caption{Four screenshots of the forsage.io website. (a) Homepage of forsage.io as of May 6th, 2021. The website is marketing the arrival of the new scheme that will be used on the Binance Smart Chain. (b) Landing page of the most profitable user showing the progress page of the X3 matrix and other macro statistics. (c) Progress page of the user's X4 matrix. (d) Progress page of the user's xGold matrix.}
\end{figure}